# What did we learn from forty years of research on semantic interference? A Bayesian meta-analysis

Audrey Bürki*, Shereen Elbuy, Sylvain Madec & Shravan Vasishth

University of Potsdam, Karl-Liebknecht-Straße 24-25, 14476 Potsdam, Germany

buerki@uni-potsdam.de, elbuy@uni-potsdam.de, madec@uni-potsdam.de, vasishth.shravan@gmail.com

*corresponding author

## Abstract

When participants in an experiment have to name pictures while ignoring distractor words superimposed on the picture or presented auditorily (i.e., picture-word interference paradigm), they take more time when the word to be named (or target) and distractor words are from the same semantic category (e.g., cat-dog). This experimental effect is known as the semantic interference effect, and is probably one of the most studied in the language production literature. The functional origin of the effect and the exact conditions in which it occurs are however still debated. Since Lupker (1979) reported the effect in the first response time experiment about 40 years ago, more than 300 similar experiments have been conducted. The semantic interference effect was replicated in many experiments, but several studies also reported the absence of an effect in a subset of experimental conditions. The aim of the present study is to provide a comprehensive theoretical review of the existing evidence to date and several Bayesian meta-analyses and meta-regressions to determine the size of the effect and explore the experimental conditions in which the effect surfaces. The results are discussed in the light of current debates about the functional origin of the semantic interference effect and its implications for our understanding of the language production system.

**What did we learn from forty years of research on semantic interference? A Bayesian meta-analysis**

**Introduction**

Most humans speak fluently, and do so without apparent effort. Yet, speaking is a highly complex skill. In the last forty years, a great deal of effort has been invested in trying to understand and model the cognitive processes and representations recruited during language production. The picture-word interference paradigm has played a key role in this endeavor. In this paradigm, participants are asked to name a picture while ignoring a distractor word, either printed on the picture or presented auditorily. The dependent measure is the response latency, or time interval between the presentation of the picture and the onset of the vocal response. A search on *Web of Science* revealed 541 articles listed under the key word *picture-word interference* between 1999 and 2018, with a total number of citations (excluding self-citations) of 18271*.* The picture-word interference paradigm has shaped our current knowledge about language production. Data obtained with this paradigm are regularly used to model word production processes and their time course (e.g., Abdel Rahman & Aristei, 2010; Levelt, Roelofs, & Meyer, 1999; Mahon, Costa, Peterson, Vargas, & Caramazza, 2007; Meyer, 1996; Roelofs, 1992; Schriefers, Meyer, & Levelt, 1990). Several experimental effects have been reported in this literature. Distractor words slow down processing (as opposed to a line of Xs, or a sequence of non-meaningful symbols, e.g., La Heij & Vermeij, 1987; Lupker, 1982). Distractor words slow down processing more than pseudowords (e.g., Klein, 1964), and among the words, low frequency words more than high frequency words (Dhooge & Hartsuiker, 2010; Finocchiaro & Navarrete, 2013; Miozzo & Caramazza, 2003; Starreveld, La Heij, & Verdonschot, 2013). The nature of the relationship between the distractor word and the word to be produced (or "target") further influences naming times. When the target and distractor are semantic associates (e.g., Alario, Segui, & Ferrand, 2000; Costa, Alario, &

Caramazza, 2005 but see Sailor & Brooks, 2014) or overlap in their spelling and /or phonology (Bi, Xu, & Caramazza, 2009; Starreveld & La Heij, 1996) the preparation of the response tends to be facilitated. By contrast, distractors of the same semantic category as the target word (target = *banana*, distractor = *apple*) tend to slow down naming times when compared with unrelated distractors. The present study focuses on this latter effect, commonly labelled the semantic interference effect.

To the best of our knowledge, the first study in which this manipulation is reported is that of Rosinski (1977). Rosinski asked participants to name lists of pictures with superimposed written distractors. He observed that participants needed more time to name a list of pictures with superimposed distractors of the same semantic category than to name the same list of pictures with unrelated distractors. Two years later, Stephen J. Lupker conducted the first reaction time study (i.e., measuring speech onset latency for each trial) comparing trials with unrelated distractors and trials with distractors of the same semantic category. He found a 30 ms difference between the two conditions (with a standard error of about 15 ms according to our estimations), again, with slower naming times for related trials. This effect was replicated many times since Lupker's study. Among the 541 studies listed under the key word *picture-word interference* on Web of Science, 316 also have *semantic interference* as a key word, with between 10 and 30 studies published per year in the last decade (see Appendix 1).

Why this enthusiasm? As we will see, many authors consider that the semantic interference effect offers a unique window into the cognitive processes engaged in language production and use this effect as a marker of underlying cognitive processes in their experiments. Crucially, however, the interpretation of these studies' empirical findings depends on the assumed functional origin (or locus) of the semantic interference effect and this locus is highly debated. The factors that modulate the effect are central to this debate and many experiments have been conducted to determine the experimental settings in which the effect arises.

Despite the fair amount of empirical data collected in the last 40 years, the picture is still unclear and many questions remain. In the present study, we present a comprehensive theoretical review of the semantic interference effect and perform a series of Bayesian meta-analyses and meta-regressions of this effect, taking into account the relevant evidence collected to date. The first aim of this contribution is to quantify the size of the semantic interference effect and the uncertainty associated with it. This information is useful to estimate the a posteriori power of existing studies and better understand past research (see Gelman & Carlin, 2014). Whereas the semantic interference effect has been reported many times, a non-negligible number of studies reported null effects (or interactions) in at least a subset of conditions. A subset of studies on the semantic interference effect was conducted in the nineties, when it was common to test a limited number of participants in many different conditions. Estimates of effect size and their precision can inform us about the likelihood that some of these studies failed to detect the effect as a result of lack of sufficient power.[1] Moreover, estimates of effect sizes can be used to design sufficiently powered experiments in future work.

The second aim of this contribution is to review and explore further the conditions in which the semantic interference effect is present. Claims regarding the factors that modulate the semantic interference effect are based on studies in which the effect is compared across conditions (e.g., with and without familiarization with the experimental material, e.g., Gauvin, Jonen, Choi, McMahon, & de Zubicaray, 2018) or on the observation that the semantic interference is present in some studies and absent in others (e.g., influence of target set size and of the repetition of this set, Roelofs, 2001). Often however, these claims do not hold when considering all published studies (for reports of semantic interference without familiarization, see Rizio, Moyer, & Diaz, 2017). Later in this section, we describe more than ten factors that have been said to modulate the semantic interference effect. With a meta-analysis, it is possible to bring in the model predictors whose dimensions vary across studies, irrespective of whether the factors were manipulated or tested by the authors of the study. For instance, the claim that familiarization matters can be tested by examining how the size of the semantic interference effect differs between experiments with and experiments without a familiarization phase.

---

[1] Estimations of effect sizes from meta-analyses are not perfect. First, they are biased by the "file drawer effect": scientific journals usually publish significant results, experiments with non-significant outcomes are less likely to be released. Moreover, when such estimates come from studies with low power, they have a higher probability of being overestimated (type M error, see Gelman & Carlin, 2014). In the present meta-analysis we were able to include a small number of unpublished datasets, some of them from the reproducibility project (Nosek et al., 2015). Moreover, in many published studies, the semantic interference effect was tested among other effects and null results are reported for a subset of experiments or conditions (e.g., Abdel Rahman & Melinger, 2009).

In the next section of this paper, we present a qualitative review of the semantic interference effect. In this review, we summarize the existing accounts of this effect, their implications for the modeling of the language production system, and discuss the factors that have been argued to modulate this effect in previous work. We then present a series of meta-analyses and meta-regressions involving the semantic interference effect. These analyses allow us to determine the size of this effect, to assess the uncertainty associated with it, and to explore the factors that interact with this effect.

**Qualitative literature review**

The literature on the semantic interference effect is tightly linked to descriptions and models of lexical access, i.e., the process by which a word's corresponding representation in long term memory is selected during language production tasks (Caramazza, 1997; Dell, 1986; Goldrick, 2006; Levelt et al., 1999). Several influential models assume that lexical selection is a competitive process, that is, the time required to select a given lexical representation is a function of the activation of other lexical representations (e.g., Abdel Rahman & Aristei, 2010; Abdel Rahman & Melinger, 2009; Belke, Brysbaert, Meyer, & Ghyselinck, 2005; Damian & Bowers, 2003; Levelt et al., 1999). In these so-called "lexical-competition accounts" the selection of a word's lexical representation occurs when the activation of this representation exceeds the activation of all other representations by a certain amount. According to other accounts (see for instance Mahon et al., 2007) lexical selection is not a competitive process. The target representation is selected when it reaches a certain activation level, irrespective of the activation levels of other representations.

Crucially for our purposes here, the major empirical argument taken to support the competitive view of lexical access is the semantic interference effect. Importantly, however, this argument only holds if the semantic interference effect indeed originates during lexical access. The functional origin (or locus) of this effect has been at the center of intense debates in the last two decades.

According to a first view, the semantic interference effect arises during lexical access, and reflects the competitive nature of this process (e.g., Roelofs, 1992; Levelt et al., 1999). During a picture-word interference task, the distractor word is automatically processed, and its corresponding lexical representation becomes activated. This activation delays the selection of the target word representation. Moreover, distractors of the same semantic category receive additional activation via the processing of the picture. Processing a picture results in the activation of the corresponding concept. This activation spreads to semantically related concepts, which in turn send activation to their corresponding lexical representations. As a result, the target word lexical representation becomes activated but semantically related representations, among which, the distractor word representation, also receive some activation (e.g., Bloem & La Heij, 2003; Glaser & Glaser, 1989; Roelofs, 1992). Because related distractors are more activated than unrelated distractors, they will compete more strongly with the target word representation. The swinging lexical network account by Abdel Rahman and Melinger (Abdel Rahman & Melinger, 2009, see Abdel Rahman & Melinger, 2019 for a recent update) is a variant of the above model. In this account, distractors of the same semantic category produce facilitation during the conceptualisation process. This is because they activate their "category node", e.g., animal, which in turns activate other lexical representations under this category node (i.e., all animals). The same distractors produce interference during lexical selection. Facilitation dominates when only few related items are activated, and competition dominates when a larger network of interrelated competitors is activated.

According to an alternative view, which builds on Lupker (1979)'s original proposal, the semantic interference effect arises during response execution. Mahon et al. (2007) provide for instance a long list of arguments and specifications for what they refer to as the "response-exclusion account". In this account, distractor words are prepared for production and stored in an articulatory buffer. In order for the target word to be articulated, the distractor must first be excluded from this buffer. Critically, the time to suppress the non-relevant response depends on its relevance. Distractors that are relevant given the semantic context are more difficult to suppress than distractors that are less relevant. In this account, the semantic interference effect occurs after lexical access. Adherents of the response-exclusion hypothesis argue that the semantic interference effect cannot be taken as empirical support in favor of the competitive nature of lexical access because it originates after lexical access.

A few authors have also suggested that the semantic interference effect could originate during pre-response selection (i.e., perceptual encoding or activation of conceptual information, Dell'Acqua et al. 2007), but the empirical arguments are scarce (see Kleinman, 2013; Piai & Roelofs 2013; Piai, Roelofs, & Roete, 2015, or Schnur & Martin, 2012). Moreover, as argued by Schnur and Martin (2012), competition at the conceptual level is hard to reconcile with the observation that other types of semantic relationships generate facilitation, or with the observation that the semantic interference effect is only found when an overt naming response is required.

It has been regularly assumed that accounts of the semantic interference effect make different predictions regarding the conditions in which the effect should be observed. The conditions under which the effect is observed (or might reverse to a facilitative effect) have therefore been examined in many studies. These are described in the next section and will be explored later in the meta-analysis.

***Boundary conditions of semantic interference in the picture-word interference paradigm***

Whereas many studies reported semantic interference effects in the picture-word interference paradigm, a non-negligible number of studies reported null effects or interactions between the semantic manipulation and other factors. Some of these factors are assumed to have important theoretical implications for the understanding of the semantic interference effect and in turn, the nature of lexical access during language production. A subset of these factors have been tested many times, and their interaction with semantic relatedness appears robust. The role of other factors is less certain, these factors having been manipulated in only a subset of studies, or their role determined a posteriori, to explain the absence of semantic interference in a subset of experiments. An overview of the factors discussed in the literature is provided in Figure 1. In the figure, the factors are grouped according to whether they concern properties of the design or of the materials, and for the latter, whether they concern the target words, the distractor words, or their relationship. In this section, we discuss each of these factors in turn, following their order of occurrence in the figure (from top left to bottom right), except for the factors *Number of semantic categories, Number of target words, and Distractors in response set,* which we discuss together with the factor *Target Repetition.*

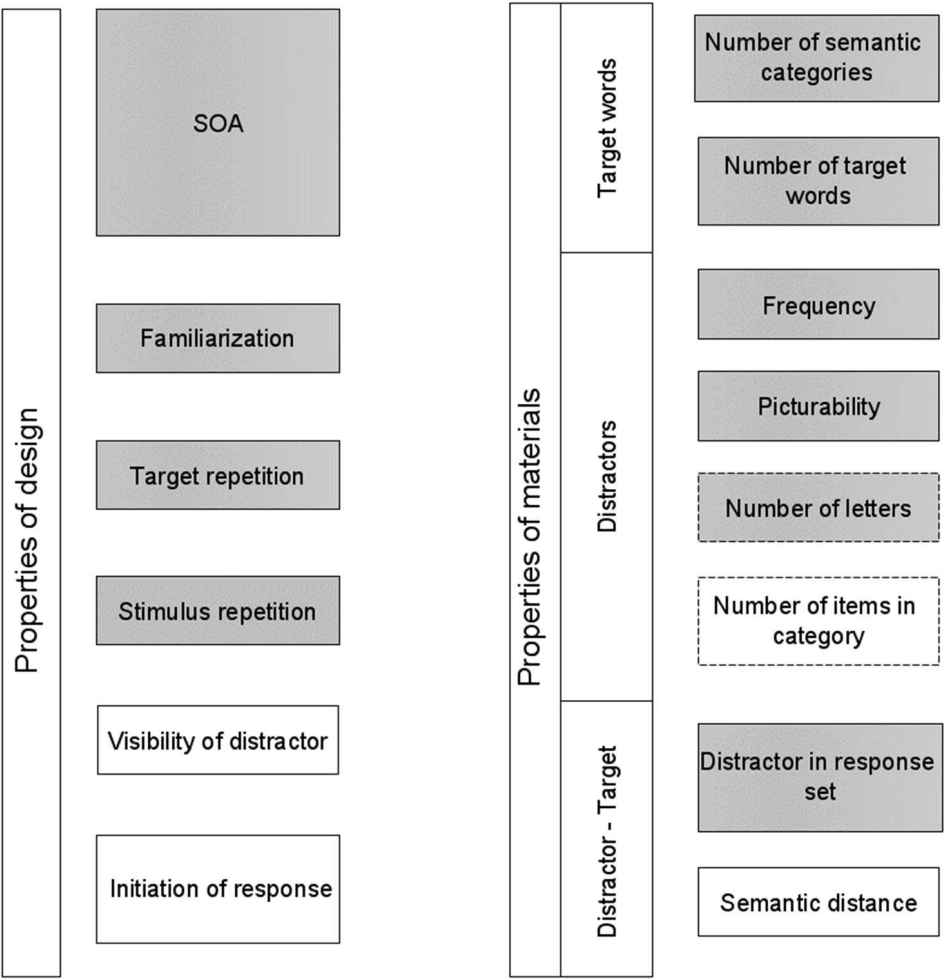

*Figure 1. Moderators of the semantic interference effect according to the literature (boxes with solid lines), or factors whose influence on the semantic interference effect is predicted by one or the other account of the semantic interference effect (boxes with dashed lines). All factors except for the factor "Number of items in category" are discussed below. This factor will be discussed in the General Discussion. Grey boxes correspond to the variables investigated in our analyses.*

***Stimulus-onset asynchronies (i.e., SOA).*** The first factor relates to the relative timing of distractor and picture presentation. Stimulus onset asynchrony stands for the time interval between the onset of picture presentation and the onset of the distractor presentation. An SOA of 0 means that both the distractor and the picture are displayed at the same time, a negative SOA means that the distractor precedes the picture, and a positive SOA means that the picture precedes the distractor. Several studies examined the interaction between SOA and semantic relatedness. The semantic interference effect is regularly found when the distractor is presented at the same time as the picture or shortly before/after (+/- 150 ms). There is however some variation in the exact SOA at which the effect is found/absent. To give just two examples, Glaser and Düngelhoff (1984) compared trials with categorically related and unrelated distractors at SOAs ranging from -400 to 400 ms with steps of 100 ms and reported a significant semantic interference effect at -100ms, 0, and +100 ms. Starreveld and La Heij (1996) tested SOAs of -200, -100, 0, 100 and 200 ms and reported semantic interference at -100 ms and 0 ms. Both these studies used written distractors (see Damian & Martin, 1999, for evidence suggesting that the SOA at which the effect surfaces depends on the modality of the distractor or on the length of distractor presentation, for other examples of interactions between SOA and semantic interference, see Damian & Bowers, 2003; La Heij, Dirkx, & Kramer, 1990, or Schriefers et al., 1990).

A few studies tested very long negative SOAs and reported facilitation rather than interference. For instance, Zhang, Feng, Zhu, & Wang (2016) found interference at -100 and 0 and facilitation at -400, -600, -800, and -1000. Python, Fargier, and Laganaro (2018) also obtained facilitation with an SOA of -700 ms. Zhang et al. (2016) argue that interference results from the activation at the lexical level, whereas facilitation results from activation at the conceptual level. Facilitation is found at long SOAs because activation decays at a faster rate at the lexical level. By the time the picture is presented and its name encoded, the activation generated by the distractor at the lexical level has already decayed while the activation at the conceptual level has not yet decayed. By contrast, Python et al. (2018) associate the semantic facilitation at long negative SOAs with post-lexical processes. It is worth noting that substantially long negative SOAs might allow speakers to consciously predict the picture name from the semantically related distractors, resulting in semantic facilitation (e.g., Python et al., 2018).

The interaction between semantic relatedness and SOA has been related to the relative time course of distractor processing and target word preparation. In lexical-competition accounts of the semantic interference effect, the activation of the semantic information associated with the distractor interferes with the preparation of the target word if and only if it coincides with the encoding step of the target word where this information is relevant, i.e., lexical access. If the distractor is presented too early, the activation of the semantic information related to the distractor has already decreased by the time the target semantic information is encoded and if the distractor is presented too late, this information can no longer interfere. The response-exclusion account can also explain the interaction between SOA and semantic relatedness by assuming that the distractor only creates interference if its phonological/phonetic form enters the response buffer before the encoding of the target word reaches a certain stage, e.g., phonological encoding. Accordingly, if by the time the target word engages in the phonological encoding process the buffer is still empty, it can no longer be filled with the distractor word. In principle, predictions regarding the SOAs at which the semantic interference effect should be observed differ between these two accounts with an effect disappearing at earlier SOAs for the lexical competition account than for the response-exclusion account. However, these predictions cannot be contrasted without knowing exactly when lexical access for the target word occurs, information that we do not have.

Notably, in both the lexical-competition and response-exclusion accounts, the relevant variable to determine whether there will be interference or not is not so much the SOA, but the relative speed with which distractor word and target word are processed/encoded. Accordingly, the speed with which participants encode the picture name (which is known to vary greatly from picture to picture and even more across participants) and the speed with which they process the distractor (also known to vary across items and participants) should modulate the semantic interference effect at constant SOAs (e.g., 0). As a consequence, the exact SOA at which a semantic interference effect is observed can be expected to vary across studies (as these studies have different sets of participants and words). To our knowledge, the interaction between semantic relatedness and naming times has not been investigated (but see Piai, Roelofs & Schriefers, 2011; 2012; Roelofs & Piai, 2017 or Scaltritti, Navarrete, & Peressotti, 2015 for distribution analyses of the semantic interference effect). Note that if the semantic interference effect is modulated by the relative speed with which distractor word and target word are processed/encoded, all the variables that influence the speed with which the distractors (or target words) are processed can be expected to modulate the semantic interference effect.

*Familiarization*. In picture naming experiments, whether pictures are presented with distractors or without, it is common practice to start the experimental session with a familiarization phase, during which the participants are presented with all the pictures used in the experiment and their names. A few authors have argued that this familiarization phase plays a crucial role in the semantic interference effect. Roelofs (1992), following Collins and Loftus (1975) assumes that when activation flows in the system, it leaves activation tags at each lemma node, with information regarding the source of the activation. During the familiarization phase, lemmas that correspond to the pictures are flagged as "permitted responses". To be selected for production in a subsequent naming phase, a lemma node must satisfy two criteria: have one such flag and an activation level that exceeds that of other lemmas. It follows (according to Roelofs, 1992) that semantic interference effects should only occur when the participants have first been familiarized with the pictures and the distractors are part of the response set (i.e., the same words are used as targets and distractors, see below) or can be indirectly activated by target words of the same semantic category, e.g., the distractor CAT is co-activated by the target word DOG via the category node animals (Roelofs, 1992; 2001).

To our knowledge, Collina, Tabossi, and Simone (2013) were the first to examine the role of familiarization empirically. They used a between-subject design and observed a significant interaction, the group that was familiarized with the pictures showed a semantic interference effect of 24 ms, the group that was not familiarized showed a facilitation effect of 25 ms. In this study, the distractors were not part of the response set. Gauvin et al. (2018) set out to replicate the interaction between familiarization and semantic relatedness. In their first experiment, where familiarization was manipulated across participants and distractors were part of the response set, they found an interaction between the two factors. In the second experiment where familiarization was a within group factor and the distractors were not part of the response set, they found a main effect of semantic interference but no significant interaction with familiarization.

Additional evidence supporting the role of familiarization is clearly needed. Whereas most picture-word interference studies did include a familiarization phase in their design, not all of them did. It must also be noted that the inclusion of a familiarization phase likely decreases the variance in the participants' responses at test. More variance can be expected in experiments/groups without familiarization. As a consequence, the number of data points required to obtain semantic interference effects without a familiarization phase, if such effects occur in this context, is likely higher than in studies with a familiarization phase.

***Number of semantic categories, Number of target words, Repetition, and Distractors in response set.***
The roles of the number of different semantic categories in the experimental set, the number of target words (i.e., set size), whether the distractor words are part of the response set (i.e., are also produced as targets) or the number of repetitions of the target words in the experiment have been the focus of intense debates between Roelofs (e.g., 2001), and Caramazza and Costa (2000; 2001). Roelofs (1992) argues that in the picture-word interference paradigm, distractors of the same semantic category which are not part of the response set should induce facilitation rather than interference. This is because they are not flagged as a possible response and can prime the target word rather than compete with it (as later noted by Roelofs, 2003, there is evidence in the Stroop literature that when the distractors are not part of the target set -the colors to be named-, the interference is weaker, Glaser & Glaser, 1989; Klein, 1986, see also Lamers, Roelofs, & Rabeling-Keus, 2010). Roelofs (1992) observed that at negative SOAs, distractors of the same semantic category that were not part of the response set showed facilitation, whereas at 0 and short positive SOAs, there was no semantic interference effect (note however that the semantically related condition included distractors of the same semantic category, but also hypernyms and hyponyms). The author discusses these results in the light of La Heij (1988)'s study, where semantic relatedness did not depend on whether the distractors were part of the response set or not (no interaction between the two factors, see also Lupker, 1979). Roelofs notes that the difference between his results and that of La Heij (1988) may be due to the fact that in his study, the response set (target words) only included one member per semantic category, whereas in La Heij's study, the response set had three members per category. According to Roelofs, distractors that are not part of the response set can indirectly be processed as response set members in designs with several target words per category. This is because naming a target word activates members of the same semantic category among which are some of the distractors.

Caramazza and Costa (2000) tested the predictions of Roelofs' hypothesis. In a first experiment they examined whether semantic interference effects could be found with distractors that were not part of the response set, and with only one member of each semantic category in the target set. They reported a semantic interference effect of about 30 ms at SOAs of 0 and -100. In another experiment (their Experiment 3), they failed to find evidence that the magnitude of the interference effect differed depending on whether the distractors were part of the response set or not. According to the authors, this result speaks against Roelofs account, which predicts greater interference for distractors that are part of the response set.

Roelofs (2001) in a reply to this argument, in turn argued that these results can be explained by assuming that "*a response set is only marked in memory when the number of responses is small and can be kept in short-term memory (…) and that response repetition helps establishing a response set in memory*" (p.284, see also La Heij & Hof, 1995, on the role of response set). In reply to this argument, Caramazza and Costa (2001) presented data in which the semantic interference was found with 11 targets, all of a different semantic category, irrespective of whether the target set had been repeated twice or ten times.

Data relevant to this discussion were also presented in Piai et al. (2012). These authors found semantic facilitation in a first experiment when the target set involved only one member per semantic category and the distractors were not part of the response set. By contrast, they observed interference in a second experiment with four members per category and with distractors part of the response set. Note that unlike in previous studies, the distractor was only presented briefly, right before the appearance of the picture. The authors introduce the *competition-threshold hypothesis* to explain these results. According to this hypothesis, distractors will only act as competitors when they reach a certain activation level. If they do not compete for selection, they can facilitate the production of the target word. Response set membership and number of members of a given semantic category in the experimental set are assumed to influence the activation level of distractors.

The paragraphs above suggest a complex picture and the available evidence does not allow robust conclusions regarding which of these factors or factor combinations modulate the semantic interference effect. Absence of interactions are often taken as evidence against interactions, and differences across studies are taken as supporting evidence of differences, without statistical tests. Clearly, more studies are needed to determine which aspects of the design/material modulate the interference effect. Moreover, even though many factors are implicated, tests that combine a conjunction of these factors are rare. A meta-analysis can help clarify some of these issues to the extent that these variables can be included in the statistical model. Note however that a subset of the explanations described above imply complex interactions between several variables. It is clear that determining the exact role of these variables and their interactions will require additional experiments in which these variables are manipulated orthogonally. It should be noted also that the implementation, in an experimental context, of some of the factors discussed above is more complex than the literature suggests. For instance, whereas the notion of "semantic category" is conceptually very simple, in practice, deciding to which category a given word belongs may be extremely difficult. For instance, should the scissors be in the category TOOLS, SCHOOL MATERIALS, KITCHEN UTENSILS?

Are the objects of a pencil case or vanity case part of a semantic category or do they share an associative relationship?

***Stimulus Repetition.*** Lupker noted (post-hoc observation) that the semantic effect was modulated by the number of times a given target-distractor pair was encountered. He argued that the semantic interference effect might be reduced the second time the target-distractor pair is presented. Along the same lines, Stroop (1935) noted that interference effects decrease with practice. Miozzo and Caramazza (2003) in their Experiment 6, examined whether semantic interference also decreased with repetition. Each picture-distractor pair was presented five times. They reported that although overall the semantic interference effect was significant, it decreased with repetition. The effect was of 69 ms for the first presentation, 51 ms for the second, 27 ms for the third, and of about 20 ms in the last two repetitions. The theoretical implications of this finding are not clear, but its methodological implications are clear. If the size of the semantic interference effect decreases with each repetition, including multiple repetitions of the same target-distractor word pair might not be the best option to increase the power of an experiment.

***Visibility of distractor (or presentation mode).*** A small subset of studies examined the effect of masking or hindering the visibility of the distractor on the semantic interference effect. Finkbeiner and Caramazza (2006) reasoned that masking the distractor should influence the semantic interference effect if this effect originates in the pre-articulatory buffer but should not influence this effect if it arises during lexical access/selection. In the framework of the response-exclusion hypothesis, masking the distractor should prevent the preparation of its phonological code, hence its entry in the response buffer. No semantic interference should therefore be observed. The authors further argued that in such cases, a facilitative effect could even be observed. Their reasoning builds on the observation that masked primes have been shown to produce facilitation in some tasks (e.g., categorization, see Finkbeiner, Forster, Nicol, & Nakamura, 2004). By contrast, if the semantic interference effect arises during lexical access/selection, masking the distractors could decrease their influence on naming times (hence the semantic interference effect) but should not lead to facilitation. In two experiments,

Finkbeiner and Caramazza (2006) observed semantic facilitation with masked distractors and interference with visible distractors (but did not test the interaction between semantic relatedness and mask type). Note that in post-hoc analyses, Damian and Spalek (2014) reported significant interference for visible distractors but no effect for masked distractors.

Piai et al. (2012) provided an alternative explanation of the interaction between presentation mode and semantic relatedness, compatible with a lexical locus of the semantic interference effect, in the framework of the *competition threshold hypothesis*. In this account, the distractor needs to exceed a certain level of activation to compete for selection. Masked distractors are unlikely to reach this threshold. If they fail to reach this threshold, they may produce facilitation as a result of spreading activation in the conceptual stratus.

The effect of masking the distractor on the semantic interference effect clearly needs to be assessed in more studies. The number of studies in which this variable was manipulated to date is small and most studies used unmasked distractors. This prevents us from exploring the role of visibility further in a meta-analysis.

*Initiation of response*. The next factor we discuss concerns the instructions that participants receive regarding the initiation of the vocal response. Most studies on the semantic interference effect used immediate naming tasks, in which participants are asked to produce their response as quickly as possible upon seeing the picture. A subset of studies used delayed naming tasks. In such tasks, participants see the picture, are asked to prepare their response but to withhold its execution until the presentation of a response cue. In delayed naming tasks, participants thus have time to prepare their response up to the phonetic encoding level during the delay. Any effect originating in higher processing levels that can be observed in the immediate version of the task should therefore vanish in a delayed task (see Laganaro & Alario, 2006). It follows that in lexical-competitive accounts of the semantic interference effect, no semantic interference effect should be observed in delayed naming tasks. Whereas such effects should be observed under a response-exclusion account is not straightforward.

According to Janssen, Schirm, Mahon, and Caramazza (2008), if the semantic interference effect originates at the stage where the output buffer is cleared, semantic interference effects can be expected in delayed naming tasks. Note that this hypothesis requires the additional assumption that in a delayed naming task, the participants wait for the response cue before emptying their pre-articulatory buffer.

Janssen et al. (2008) tested these predictions using an immediate naming task and a modified version of the delayed naming task. In the immediate (or standard) version of the task, the picture and distractor appeared simultaneously on the screen and participants were asked to produce their response as quickly as possible. In the delayed version of the task, the distractor appeared 1000 ms after the picture and participants were instructed to produce the target upon seeing the distractor. In two experiments, the authors reported a semantic interference effect in both the immediate and delayed tasks and took this result to support the response-exclusion hypothesis. Note however that during the 1000 ms between the onset of picture presentation and the appearance of the distractor, participants have had time to put the target word in the output buffer, and it becomes difficult to explain how the distractor can then enter the buffer given the assumption that only one response can be in the buffer at a time. Moreover, no semantic interference effect was found in a replication of this task within the reproducibility project (Open Science Collaboration, 2015), or in other attempts (Mädebach, Oppermann, Hantsch, Curda, & Jescheniak, 2011; Piai et al., 2011). These studies found semantic interference in the immediate version of the task but not in the delayed version. In sum, there is very little empirical evidence so far to support the hypothesis that semantic interference effects occur in delayed picture-word interference tasks.

***Frequency and other properties of the distractors.*** Miozzo and Caramazza (2003) observed that amongst distractors that are unrelated to the target picture, low frequency distractors interfere more than high frequency distractors. In an attempt to better understand the mechanisms underlying this distractor frequency effect, they conducted an experiment in which they manipulated both the lexical frequency of the distractor and the semantic relationship between target and distractor. They observed a semantic interference effect for both high frequency and low frequency distractor words, at -100 and 0 SOA but no effect was detected at +100 SOA. By contrast the distractor frequency effect was found at all SOAs. The authors took this result to suggest that the two effects have different loci, and are independent.

The interaction between lexical frequency and semantic relatedness was examined in subsequent studies, and used as a test case to disentangle the response-exclusion and lexical-competition accounts of the semantic interference effect. According to Starreveld et al.'s (2013), in models where lexical access is competitive and in which the semantic interference effect originates in lexical access, the two effects (i.e., distractor frequency and semantic category) have the same locus and can therefore interact. By contrast, in the response-exclusion hypothesis, the mechanisms underlying the two effects are assumed to be different and the effects are not expected to interact. The semantic interference effect occurs because it takes more time to empty the buffer when the distractor in the buffer is a relevant response. High and low frequency distractors do not differ in their response relevance and should not interact with the semantic interference effect (see Starreveld et al., 2013, e.g., p. 639). In our opinion, however, an interaction can still be expected in this account, under the assumption that when a distractor is processed too late, it can no longer enter the buffer. Given that low frequency distractors are processed more slowly, the probability that they miss the time window to enter the response buffer is higher for them than for high frequency words. Notably, under the response-exclusion hypothesis, all the variables that influence the speed with which the distractor is processed can potentially modulate the semantic interference effect. This prediction can be tested by looking at other properties of the distractors known to influence reading speed. One such property is word length. Word length has been shown to modulate processing times in several reading paradigms (e.g., Barton, Hanif, Björnström, & Hills, 2014; or Schuster, Hawelka, Hutzler, Kronbichler, & Richlan, 2016).

Starreveld et al. (2013) reported an interaction between frequency and semantic relatedness, with a smaller interference effect for high frequency distractors. Finocchiaro and Navarrete (2013) manipulated distractor frequency and semantic relationship between target and distractor but did not report results on their interaction. More evidence in favor of this interaction or its absence is clearly needed and can easily be obtained in a meta-analysis provided that the distractors or their frequencies are available.

Finally, Lupker (1979) reasoned that if the semantic interference effect arises because the distractor word must be suppressed so that the target word can enter the articulatory buffer, the suppression process should be easier when the distractor word is not relevant for the task. He further reasoned that one aspect of the words that make them relevant for the picture naming paradigm is their picturability. According to the response-exclusion hypothesis, distractors that are not picturable should produce less interference. He tested this prediction (his Experiment 4) and indeed found more interference from picturable distractors.

***Semantic distance.*** The semantic distance between two words refers to the degree to which the two words are semantically associated. Words of different semantic categories can be more or less distant (*mouse* and *cheese* are closer than *mouse* and *milk*). Likewise, words of the same semantic category can be more or less distant (e.g., *zebra* and *horse* are closer than *mouse* and *zebra*). The interaction between semantic distance between target and distractor words and semantic relatedness (same vs. different semantic categories) is seen as a crucial test case for disambiguating different accounts of semantic interference. A core assumption of competitive models of lexical access is that the time to name a given word increases with increased activation levels of the word's competitors. In such accounts, it is usually assumed that when participants name a picture in the context of a distractor word, the amount of activation that propagates from the conceptual to the lexical level increases with an increase in semantic similarity between the two words (e.g., Caramazza, 1997; Starreveld & La Heij, 1996). This assumption predicts an interaction between semantic relatedness and the semantic distance between target and distractor words, with greater interference from distractors that are semantically close to the target words (see also Mahon et al., 2007). The response-exclusion account predicts no such modulation of the interference effect.

Abdel Rahman and Melinger (2009)'s swinging network model makes more nuanced predictions. In this account, semantic association creates facilitation at the conceptual level, while category membership creates competition at the lexical level. Close distractors should lead to more facilitation than far distractors at the conceptual level. Moreover, semantically far distractor-word pairs (e.g., *bee-horse*) are expected to co-activate a larger cohort than closely related distractor-word pairs (e.g., *bee-ant*), as a result the latter should induce less interference.

Several studies examined the interaction between semantic distance and semantic relatedness, the resulting picture is once again heterogeneous (see **Appendix 2** for a detailed summary), with studies reporting no interaction between semantic distance and semantic interference (Hutson & Damian, 2014; Lupker, 1979), studies reporting changes in the direction of the effect of semantic relatedness at different SOAs for weakly and strongly associated target-distractor pairs (e.g., La Heij et al., 1990, Experiment 2), studies reporting a decrease in semantic interference with decreasing semantic association between distractor and target (Vigliocco, Vinson, Lewis, & Garrett, 2004) or a reversed pattern, that is interference for weakly associated target-distractor pairs only (i.e., *whale* produced interference for the word *horse*, but *zebra* did not, Mahon et al., 2007).

The issue of whether semantic similarity truly modulates the semantic interference effect also has potential methodological implications. If semantic distance modulates the interference effect, the question arises of how much of the differences across studies can be attributed to differences in semantic distance as opposed to other factors. Semantic distance was not controlled for in many studies and the material used shows substantial variability in this dimension. Interestingly, Aristei and Abdel Rahman (2013) found that the semantic interference effect was no longer significant in their data once they introduced semantic similarity between distractor and target in their analysis.

*Factors influencing semantic interference effect: Summary*

The literature provides robust evidence that the semantic interference effect is modulated by the SOA, or time interval between the presentation of the picture and that of the distractor. When a semantic interference effect is reported, the distractor was either presented slightly before the picture, slightly after, or at the same time. Interactions between SOA and semantic relatedness are reported in many studies but the exact SOAs at which the effect surfaces varies across studies. For all the other factors discussed in the literature, the evidence is far from homogeneous and no strong conclusions about the relevance of these factors is warranted. For many of these factors, the existing empirical evidence is unconvincing, either because the interaction between this factor and semantic relatedness was reported in one or two studies only, because the interaction was found in some studies but not in others, because the evidence relies on the absence of an effect, or because some studies showed a semantic interference where the interaction in other studies suggest there should be none. It is worth noting also that a fair number of the designs reviewed above are likely underpowered. As a consequence, it is possible that effects or interactions went undetected (type II errors) and that others were overestimated (type M errors). Moreover, most studies tested one or at best two modulating factors at once, without controlling for others. As already discussed by many authors, it is likely that several factors combine to modulate the semantic interference effect. Additional studies that manipulate the factors and test their joined contributions are clearly needed.

Importantly, decisions to consider a factor as a modulator of the semantic interference effect in this literature are all based on significance. Claims regarding the existence or absence of an effect based on significance alone can be misleading (see Nicenboim, Roettger, and Vasishth 2018 for detailed explanations and arguments). This is because the outcome of a single study is influenced by the sample tested, the method used, and the choice of the statistical model. Moreover, a *p*-value below 0.05 merely tells us that the probability is smaller than 0.05 of observing the estimated effect, or an estimated value that is more extreme conditional on the assumption that the effect is 0; this conditional statement does not imply that the effect is present, it merely implies that the observed effect is unlikely if we assume that the true effect was zero. Finally, a *p*-value above 0.05 cannot generally be taken to support the hypothesis that there is no effect or no interaction (e.g., Hoenig & Heisey, 2001). The accumulation of evidence across studies is much more informative regarding the robustness of a given empirical result. The purpose of a meta-analysis is precisely to synthesize the evidence across all relevant studies. It provides an estimate of the underlying effect of interest based on the available data. The contribution of each study is weighted by the precision (uncertainty) of its estimate; studies with less precision (i.e., with a larger standard error) have a smaller impact in determining the overall effect than studies that have greater precision (a small standard error).

As discussed at several places above, the conditions under which the semantic interference effect is found or is absent often feeds the debate about the locus of the effect. Our theoretical review and discussion of the factors that modulate the semantic interference effect reveal however that there are very few factors for which contrastive predictions can be made. When such predictions are made in the literature, they do not necessarily survive a closer analysis.

Moreover, the interaction between semantic interference and SOA suggests that the semantic interference effect is modulated by the relative speed with which the target and distractor words are processed/encoded. If this is true, then any variable that influences processing times for the distractor (e.g., distractor frequency) or processing times for the target word (e.g., repetition, familiarization) can be expected to interact with the semantic interference effect. Such an interaction can easily be explained by competitive and non-competitive accounts of semantic interference. Moreover, both types of accounts would predict the same type of interaction, i.e., that the semantic interference effect decreases when the processing of the distractor is delayed relative to the encoding of the target word. In lexical-competition accounts, distractor words can only interfere with the semantic interference effect if the semantic information is active during the semantic encoding of the target word. It seems reasonable to assume here that unless the distractor is presented at a negative SOA, the activation of its semantic information is unlikely to have decayed sufficiently by the time the semantic/conceptual information associated with the target word is activated. The word production literature on picture naming agrees on the assumption that the activation of the conceptual / semantic information of the picture's name occurs within the first 200 ms after picture onset (see for instance Indefrey, 2011). The activation of the meaning of a written word likely arises, at the earliest, in this same time window (see for instance Hauk, Coutout, Holden, & Chen, 2012). The semantic interference effect can therefore be expected to decrease only when distractor processing is delayed relative to the encoding of the target word. In the response-exclusion account, the distractor can only interfere with the production of the target word if it can enter the response buffer before the encoding of the target word has reached a certain point. In this account, any factor that delays the processing of the distractor relative to the encoding of the target word would result in less interference.

Finally, whereas the mechanisms underlying the semantic interference effect are radically different in the two accounts, they make similar predictions for several variables. In lexical-competition accounts, variables that enhance the activation of distractor words are expected to increase the competition, in the response-exclusion account, variables that enhance the relevance of the distractor will make it harder to be suppressed. Several variables are likely to modulate activation and relevance in the same way. Distractors are expected to be more activated and/or more relevant if they are in the response set, even more so if there is a familiarization phase. They are expected to be less activated and/or less relevant relative to the target words if the latter are repeated many times. In sum, for most variables an interaction with the semantic interference effect does not disentangle accounts of the semantic interference effect. We will come back to this in the General Discussion.

**Bayesian meta-analyses**

The studies reviewed in the introduction relied on significance to determine whether the semantic interference effect was present or not and whether its presence depended on the experimental conditions. A frequentist meta-analysis can be conducted in this framework, to determine whether the effect of interest, here the semantic interference effect, is significant when considering all previously published studies of this effect. Alternatively, such meta-analysis can be conducted in the Bayesian framework. Bayesian analyses provide information about the effect of interest by determining the probability of this effect given the data and model (see Vasishth & Nicenboim, 2016). A Bayesian analysis takes as input the data (the observed difference in means between the two conditions being compared, and the estimated standard error of this observed difference) and prior distributions for each of the model parameters. The output is a posterior distribution for each of these parameters. Posterior distributions represent the possible values of the parameters given the data and model. These posterior distributions provide information about the estimate of the effect given the data and model together with uncertainty about this estimate. Posterior distributions can be used to determine the probability of an effect being larger or smaller than zero. Probabilities are given by the area under the posterior distribution. Often, these posterior distributions are used to compute so-called credible intervals. A 95% credible interval is an interval in which we are 95% certain that the true value of the parameter lies (e.g., Vasishth, Nicenboim, Beckman, Li, & Jong Kong, 2018, see Morey, Hoekstra, Rouder, & Wagenmakers, 2016 on *Confidence* Intervals).

In the present study, the input of the analyses are the estimates of the semantic interference effect in previous studies. We are interested in knowing the probability distribution of the semantic interference effect, given the data and model. As in other linear models, it is possible to add additional predictors in the meta-analysis. This way, we can obtain estimates for these predictors and information about the probability that the influence of a given predictor on the interference effect is greater than zero given the data.

Our analyses have three aims. The first is to obtain the posterior probability of the semantic interference effect considering all the available datasets with SOAs of zero or very short SOAs around zero. This restriction is based on the widely shared assumption that the semantic interference effect is found at these SOAs, assumption in turn grounded in the many studies reporting interactions between SOA and semantic relatedness. The second aim is to provide information on the power of existing studies. The third is to explore the role of different factors on the semantic interference effect. The variables considered in the model need not have been manipulated in all studies, but must show sufficient variation across studies. In the present study, we were able to explore the roles of the following variables: SOA, naming times, distractor frequency, distractor length, distractor picturability, familiarization, whether distractors are part of the response set, number of semantic categories, set size, and target repetition (see Figure 1). Importantly, these analyses must be considered exploratory. Many of the factors were not explicitly manipulated and as a result, causality is difficult to assess. Moreover, some of the factors are unbalanced, with some levels being underrepresented (e.g., only 18 studies did not use familiarization). The outcome of these analyses should serve as a basis to implement further controlled experiments in which the factors are manipulated.

*Eligibility criteria and dataset selection*

Experiments were included in the meta-analysis if all or a subset of trials satisfied the following conditions. The task was a classical picture interference task: participants were presented with pictures of objects that they had to name, the pictures were accompanied by written or spoken distractor nouns, some of them unrelated to the picture, some of them of the same semantic category (e.g., horse-cat). We focused solely on basic level naming, that is naming an object of a given semantic category in the context of a distractor member of the same category. We did not include studies where the distractor was a subordinate (e.g., tulip-flower; Hantsch, Jescheniak, & Schriefers, 2005) or a superordinate (e.g., bird-seagull, Kuipers, La Heij, & Costa, 2006) of the target noun, where targets and distractors were involved in a part-whole relationship (e.g., window-house; Sailor & Brooks, 2014) or where targets and distractors were both subordinates (e.g., daisy-rose, see Vitkovitch & Tyrrell, 1999). In all included experiments, healthy adult participants produced a vocal response, which was either a noun (in the majority of studies) or a noun preceded by a determiner (e.g., Schriefers &Teruel, 2000). Studies in which the expected responses were longer noun phrases (e.g., Meyer, 1996) were not included. Studies in which pictures (e.g., Roelofs, 1993) or distractors were not nouns or studies involving dual tasks (e.g., Piai et al., 2015) were also not included. The meta-analysis was further restricted to experiments conducted in the participants' first language. We disregarded the few studies (e.g., Rosinski, 1977) in which only the total time to name all pictures in a list were provided and focused on those studies where the naming times were collected for each trial. Given that most studies used immediate naming tasks, we decided to focus on this task and disregarded the datasets collected with delayed naming tasks. Finally, we disregarded the few experiments in which the distractor was presented very briefly followed by a mask (e.g., Blackford, Holcomb, Grainger, & Kuperberg, 2012) or restricted the datasets to the non-masked trials (e.g., Finkbeiner & Caramazza, 2006; Piai et al., 2012).

We started the search for experiments corresponding to these criteria in our own databases of articles, and extended this search to the references cited in those articles. At the end of this process, we performed a quick search in the Mendeley database with the key words "picture-word interference" and "semantic interference", and added the studies which were returned and not yet in our database[2].

This resulted in a total of 66 journal articles with a total of 103 experiments corresponding to our selection criteria. We had to disregard one experiment for which we did not have enough information to estimate the effect size or the standard error. We also searched the OSF reproducibility project for replications of these studies and found one experiment, which we added to our database. We contacted all authors who had at least one publication between 1990 and 2018 to request their data and asked for potential unpublished data. Many responded positively and the data corresponding to 41 published experiments, one unpublished experiment and a pilot study were shared with us. Figure 2 represents the workflow leading to the final set of studies considered in the meta-analyses and their sources.

Many of the experiments included in the meta-analysis did more than just contrast semantically related and unrelated distractors. Some of them included additional distractor types, such as identity (distractor = target word), a phonologically related word, or a semantically associated word. In such cases, trials with distractors from other conditions were disregarded.

The selected experiments were further divided in a set of 198 studies. Thereafter, "study" is used to refer to an experiment or experiment part for which we extracted or computed an estimate of the semantic interference effect. Each study corresponds to an experiment or an experiment part that involved related and unrelated trials. A few experiments manipulated the level of semantic association or semantic similarity between target and distractor. That is, distractors of the same semantic category as the target word could be more or less similar or associated with the target (e.g., Vieth, McMahon, & de Zubicaray, 2014). In such cases, and whenever this was possible, we considered each level of semantic similarity as a different study. Many experiments also manipulated aspects of the design. Depending on the factor manipulated, we either computed an estimate for each level (e.g., SOA, familiarization, modality of presentation of distractor, distractors in response set or not) or pooled across levels of the other variable (target set size e.g., La Heij & Hof, 1995; La Heij & Vermeij, 1987, distractor duration; Alario et al., 2000; Piai, unpublished) [3]. Another subset of experiments manipulated the frequency of the distractors or of the target word and this manipulation was orthogonal to that of semantic relatedness (e.g., Mädebach et al., 2011; Miozzo & Caramazza, 2003). For these studies, we estimated the semantic interference effect pooling across frequency levels. Our final database included 198 studies, with raw data available for 67 of them. The list of experiments included in the meta-analysis with detailed information on the subsetting and extraction procedures can be found in the supplementary material.

---

[2] The search on Mendeley was performed in 2018. It is likely that despite our efforts, we missed a few relevant papers, especially if these have been published in a different literature and/or referenced under other key words.

---

[3] Decisions to pool across the levels of a variable or to split the experiment according to the levels of a variable were driven by a mixture of theoretical arguments, methodological arguments and/or information available in the papers. We pooled across levels of distractor frequency because it was possible to compute a lexical frequency value for each study, and test for the interaction between lexical frequency and semantic relatedness across all studies. We split experiments according to the SOA condition used, because it was possible to assign an SOA value to each study, and to enter this variable as a predictor in the analysis. Moreover, given the empirical evidence suggesting that long negative SOAs show facilitation, it would not have made much sense to estimate the semantic interference for a dataset including wide ranges of SOAs. In most cases, for factors whose levels had little variation across studies, the data were pooled across levels. However, in several cases, we had to split the data because the statistics provided in the original article were only provided for sublevels of the related and unrelated condition. For instance, Rayner and Springer (1986, Experiment 1) provided the statistics for different conditions, differing in whether the distractor preserved the first letter and/or the shape of the distractor.

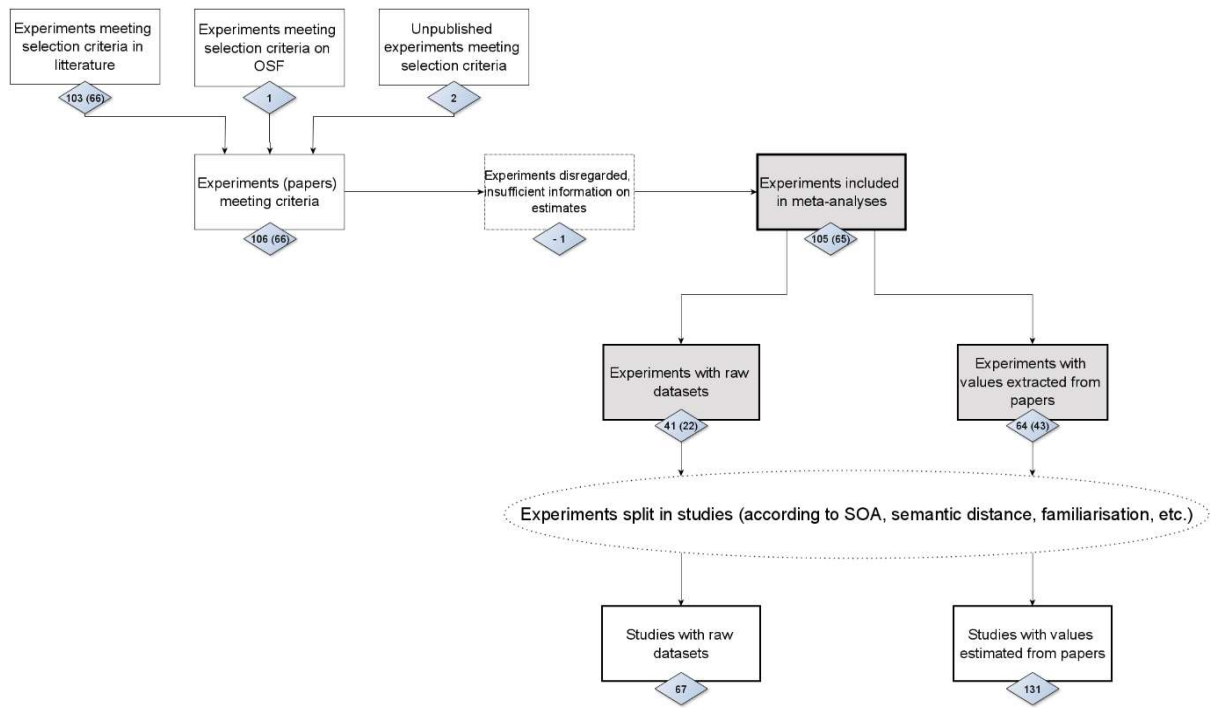

*Figure 2. Flowchart representing the selection procedure and source of the studies included in the meta-analysis. Numbers in the diamonds represent the number of experiments (or studies for the two boxes at the bottom of the figure), the numbers in brackets correspond to the numbers of different articles in which these experiments are published.*

*Extraction of estimates*

When the raw datasets were available (67 studies), we followed the procedures described by the authors to remove incorrect responses, outliers, items or participants. In a few cases, we further removed data points with response times at or below zero. We then extracted the estimates of the semantic interference effect together with the Standard Error of this estimate using a linear mixed-effects models (library lme4, Bates, Mächler, Bolker, & Walker,2015) with untransformed naming times as the dependent variable. We used effect coding to code the variable "semantic relatedness", with 0.5 for semantically related and -0.5 for unrelated distractors. This way, the estimate represents the difference between the two conditions (and a positive estimate is expected whenever there is interference), and the intercept represents the grand mean. The random effect structure of the model included intercept for item and participant and a slope for the effect of semantic relatedness for participant and item (to the extent that the item random slope was supported by the design, i.e., same target words used in related and unrelated conditions). Note that in most cases, our analysis differed from the original analysis reported in the paper, therefore, our estimates do not necessarily match the ones reported in the published analyses for these datasets. When the raw datasets were not available, we used the information provided in the published papers. Most studies provided means in the semantically related and in the semantically unrelated conditions, which allowed us to compute the mean difference in milliseconds (i.e., semantic interference effect). In two cases, this mean difference had to be estimated from a figure. The estimation of the Standard Error of this difference was rarely provided in the papers. We computed this value based on the information provided in the paper. **Appendix 3** details the information and computations that we used to estimate these Standard Errors. Information about the computation used for each individual dataset can be found in the supplementary material provided with this paper. Note that in many cases, the information available was limited and as a result, the Standard Errors are likely to be underestimated. This is for instance the case for all studies in which only by-participant analyses are reported, or when no other information than "not significant" or F< 1 is provided. In some cases, the paper did not provide enough information to compute the SE for the comparison of interest. For instance, for some datasets, we had no information to compute the SE of the estimate at a given SOA. In those cases, we used the statistics available for the analysis with all SOAs.

*Analysis*

We performed different types of analyses. Table 1 below summarizes these analyses and Figure 3 summarizes the selection of the studies included in each of these analyses. The datasets and scripts to reproduce these analyses can be found on OSF (osf.io/k6f4c/). Meta-analyses provide an estimate of the size of an effect of interest $\theta$ (e.g., difference between trials with related vs. unrelated distractors) and of its degree of precision based on a set of studies where this effect and its Standard Error can be computed or is reported. The input to the meta-analysis consists in the estimates and Standard Errors for the effect in the individual studies. In the present work, we performed a meta-analysis of the semantic interference effect (Analysis 1), and a series of meta-analyses examining the interaction of semantic interference with the properties of the distractors (Analysis 5).

In meta-regressions the effect size $\theta$ is estimated for different values of one or several predictors. Meta-regressions thus allow us to explore the role of moderating variables on this effect size. In the present work, we performed two meta-regressions. In the first, we considered all datasets and examined the role of SOA on the semantic interference effect (Analysis 3). In the second, we considered all datasets with SOAs of zero and short SOAs around zero and examined the role of the properties of the design on the semantic interference effect (Analysis 4).

Two kinds of models can be used to conduct a meta-analysis (or meta-regression); a fixed-effect or a random effect meta-analysis. A random effects meta-analysis assumes that each study has a different true effect $\theta_i$. By contrast, a fixed-effects analysis assumes the same underlying true effect $\theta$ for all studies. Here we opted for random-effects analyses. Given that the studies that we included were conducted under different conditions, with different types of materials and in different languages it makes more sense to assume a different true effect $\theta_i$ for each of them. Note that the outcome of the analysis provides information on the variance between studies.

We performed our meta-analyses under the following assumptions. Each study *i* has an underlying true mean $\theta_i$ generated from a normal distribution with a mean of $\theta$ and a variance $\tau^2 = 100^2$. The observed semantic interference effect $y_i$ in each study included in the meta-analysis is assumed to be generated from a normal distribution with mean $\theta_i$ and variance $\sigma_i^2$, the true standard error of the study. The model specifications are displayed in Equations (1). Note that this model is in fact a mixed-effects model with a random intercept for the random variable *study*. (1)

$$y_i | \theta_i, \sigma_i^2 \sim N(\theta_i, \sigma_i^2) \; i = 1, \dots, n,$$

$$\theta_i | \theta, \tau^2 \sim N(\theta, \tau^2),$$

$$\theta \sim N(0, 100^2),$$

$$\tau \sim N(0, 100), \tau > 0$$

$y_i$ is the observed semantic interference effect (i.e., difference between the related and unrelated conditions) in study *i*; $\theta$ is the true interference effect to be estimated by the model; $\sigma_i^2$ is the variance for study *i*, estimated from the standard error of the semantic interference effect for this study; and $\tau^2$ is the between-study variance.

In meta-regressions, the model has additional parameters to estimate, the regression coefficients $\beta_j$. The model specifications are displayed in Equation (2).

(2)

$$y_i | \theta_i, \beta_j, \sigma_i^2 \sim N(\theta_i + \beta_j * predictor_{i,j}, \sigma_i^2) \quad i = 1, \dots, n, j = 1, \dots p$$

$\beta \sim N(0, 100^2)$, for categorical predictors and $\beta \sim N(0, 20^2)$, for continuous predictors

Predictor is a regression predictor in the meta-regression. The number of $\beta$ parameters *p* in the meta-regression is equal to the number of predictors. In the meta-regression, $\theta_i$ is the semantic interference effect adjusted for the effects of the predictors $\beta_j$. We used normal priors with a mean of 0 and a standard deviation of 100 for the intercept, the standard deviation and the categorical predictors and normal priors with a mean of 0 and a standard deviation of 20 for continuous predictors. We also performed sensitivity analyses with different priors, these are detailed in the results section. The analyses were performed in R (R Core Team, 2018) using the package brms (Bürkner, 2018).

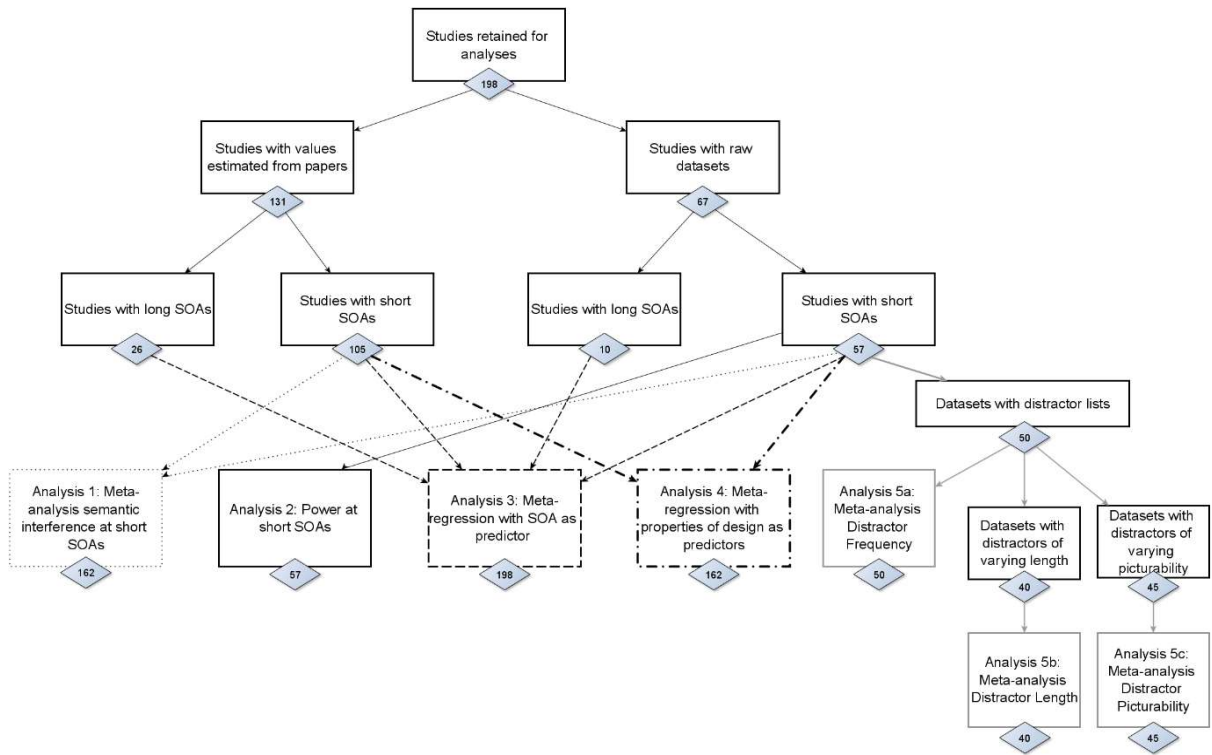

*Figure 3. Dataset selection for each analysis*

*Table 1. Summary of statistical models*

| Name | Aim | Input | Predictor(s) | Dataset |
|---|---|---|---|---|
| *Analysis 1: Meta-analysis semantic interference at short SOAs* | Provide estimate of semantic interference effect and its precision at short SOAs | • Estimates of semantic interference<br>• Standard errors of these estimates | - | 162 studies with SOAs ranging from -160 to + 150 ms |
| *Analysis 2: Power at short SOAs* | Determine a posteriori power of existing studies | • Estimates of semantic interference<br>• Standard errors of these estimates<br>• Variance random intercept and slope for items and participants<br>• Residual variance<br>• Number of participants and items used in study<br>• Meta-analytic estimate of semantic interference computed in Analysis 1 | - | 57 studies with SOAs ranging from -160 to + 150 ms whose raw data were available |
| *Analysis 3: Meta-regression with SOA as predictor* | Investigate the effect of SOA on the semantic interference effect | • Estimates of semantic interference<br>• Standard errors of these estimates | • SOA | All 198 studies |
| *Analysis 4: Meta-regression with properties of design as predictors* | Investigate the effects of the design on the semantic interference effect | • Estimates of semantic interference<br>• Standard errors of these estimates | • SOA<br>• Mean Naming times in unrelated condition (mean RT)<br>• Interaction between SOA and mean RT<br>• Number of repetition Target word<br>• Number of semantic categories in target set<br>• Familiarization<br>• Whether distractor in response set<br>• Set size | 162 studies with SOAs ranging from -160 to + 150 ms |
| *Analysis 5: Meta-analyses of the interaction between semantic interference and (a) Distractor Frequency, (b) Distractor Length and (c) Distractor Picturability* | Investigate the interactions between distractor properties and the semantic interference effect | • Estimates of interactions between semantic interference and:<br>- 5.a distractor frequency<br>- 5.b Distractor length<br>- 5.c Distractor Picturability<br>• Standard errors of these estimates | | Studies whose datasets were available with SOAs ranging from -160 to + 150 ms minus datasets with no information on distractor or no variability in distractor property |

**Results**

*Analysis 1: Meta-analysis semantic interference at short SOAs*

The purpose of the first analysis was to quantify the size of the semantic interference effect and its precision, at SOAs where the effect is commonly assumed to be found. We selected the 162 datasets with short negative SOAs (-1 to -160 ms), zero, or short positive SOAs (1 to 150 ms). Figure 4 displays the posterior distributions of the estimates of the semantic interference effect for each study, weighted by the meta-analysis. A more detailed version of this forest plot, with information of the corresponding studies for each estimate, can be found in the supplementary material.

This analysis reveals that the overall effect is about 21 ms with a 95% Credible Interval –hereafter CrI– ranging from 18 to 24 ms. The posterior distribution of the between-study standard deviation had a mean of 14.5 ms (CrI: 12.1-17.2). The posterior distributions of the semantic interference estimate and of the between-study standard deviation are plotted in Figure 5. A sensitivity analysis was conducted with two different priors for the semantic interference effect (normal distribution with a mean of 0 and a standard deviation of 200 and a uniform distribution bounded between 0 and 200). These priors implement the assumptions that values below zero or above 200 ms for the semantic interference effect are highly unlikely. The results of all analyses are displayed in Table 2.

*Table 2. Sensitivity analysis: Meta-analysis estimate and 95% credible interval for different priors*

| *Prior Intercept* | *Estimate (ms)* | *CrI (95%)* |
| --- | --- | --- |
| **N(0,100)** | **21.2** | **18.3 – 24.3** |
| N(0,200) | 21.3 | 18.6 – 24.2 |
| Uniform(0,200) | 21.2 | 18.3 – 24.2 |

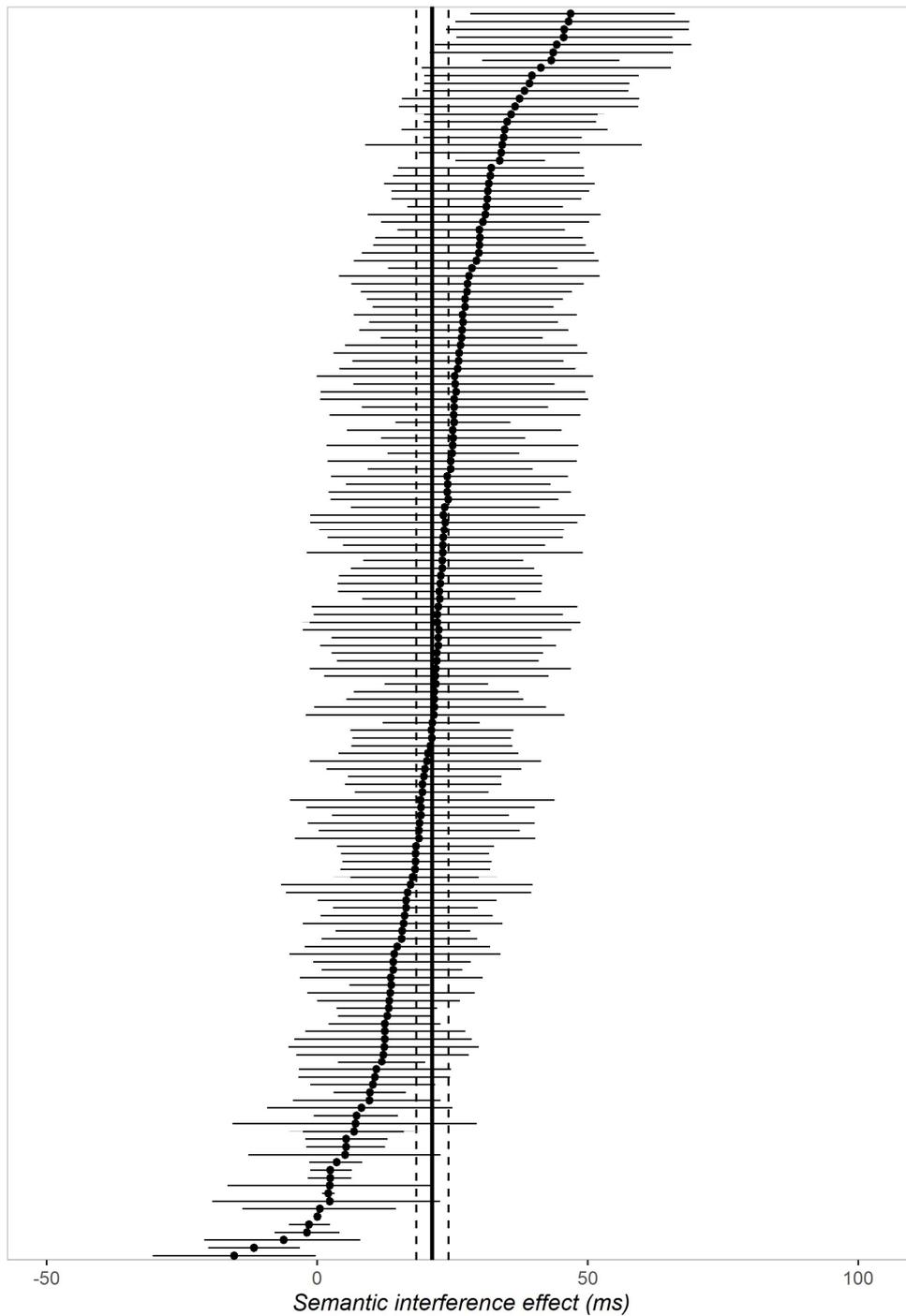

*Figure 4. Summary of the random-effects meta-analysis modeling the semantic interference effect for the 162 studies with short SOAs. For each study, the figure displays the posterior estimate (mean and 95% credible interval) for the interference effect. A positive value means interference, a negative value signals facilitation. Solid and dashed vertical lines in the background mark off the grand mean (i.e., the meta-analytic effect) and its 95% credible interval. A more detailed version of this figure, with the name of the different studies can be found in the supplementary material.*

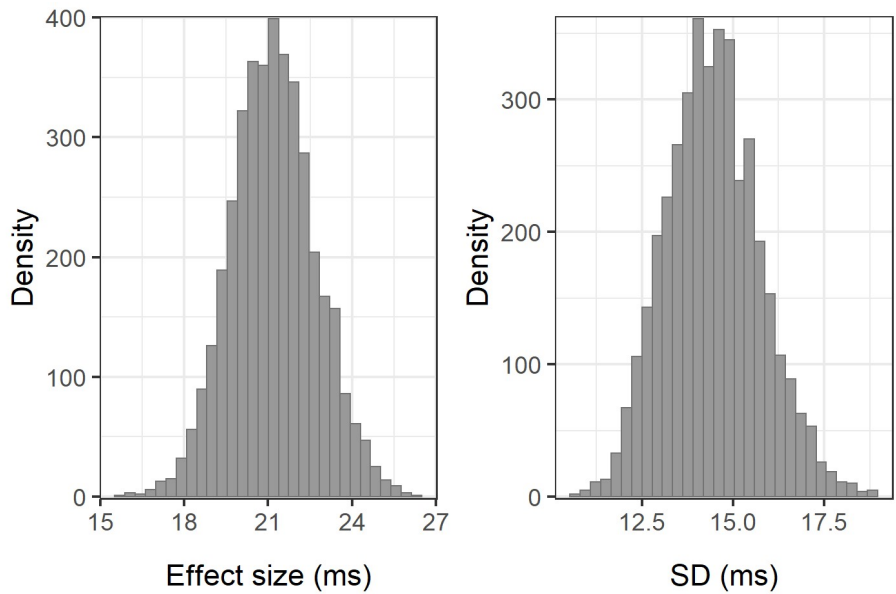

*Figure 5. Posterior distribution of the semantic interference estimate (left) and between-study standard deviation (right) in the meta-analysis of the semantic interference effect at short SOAs (N = 62)*

*Analysis 2. Power at short SOAs*

The aim of this analysis was to obtain information of the power of previous studies considering the effect size estimated from the meta-analysis and using mixed-effects models with by-participant and by-item random slopes for the experimental effect. Mixed-effects models have become the gold standard in psycholinguistic research. The inclusion of by-participant and by-item random slopes for the effect(s) of interest ensures that the effects reported generalize over participant and items (e.g., Bates et al., 2015; Pinheiro & Bates, 2000). The power of a study is partly determined by the statistical analysis. As explained in details for instance in Judd, Westfall, and Kenny (2012, see also Brysbaert & Stevens, 2018), the computation of power in the context of a mixed-effects model requires that the random part of the model be taken into account.

We restricted the set of studies to the 57 for which we had the raw datasets, disregarding those studies with SOAs large than 160 (absolute values). For each dataset, a linear mixed-effects model was fit with varying intercepts and slopes, without any correlation parameter for random effects. The mean parameter estimates for all fixed effects and variance components were recorded. Then, for each study, prospective power was computed using these parameter estimates and the meta-analysis mean, lower and upper bounds (i.e., 21 ms, with 95% credible intervals 18 and 24 ms). This led to three power distributions: for the mean, lower, and upper bound of the meta-analysis estimates. These distributions are plotted in Figure 4 below. This analysis suggests that the majority of datasets had a power below 80%.

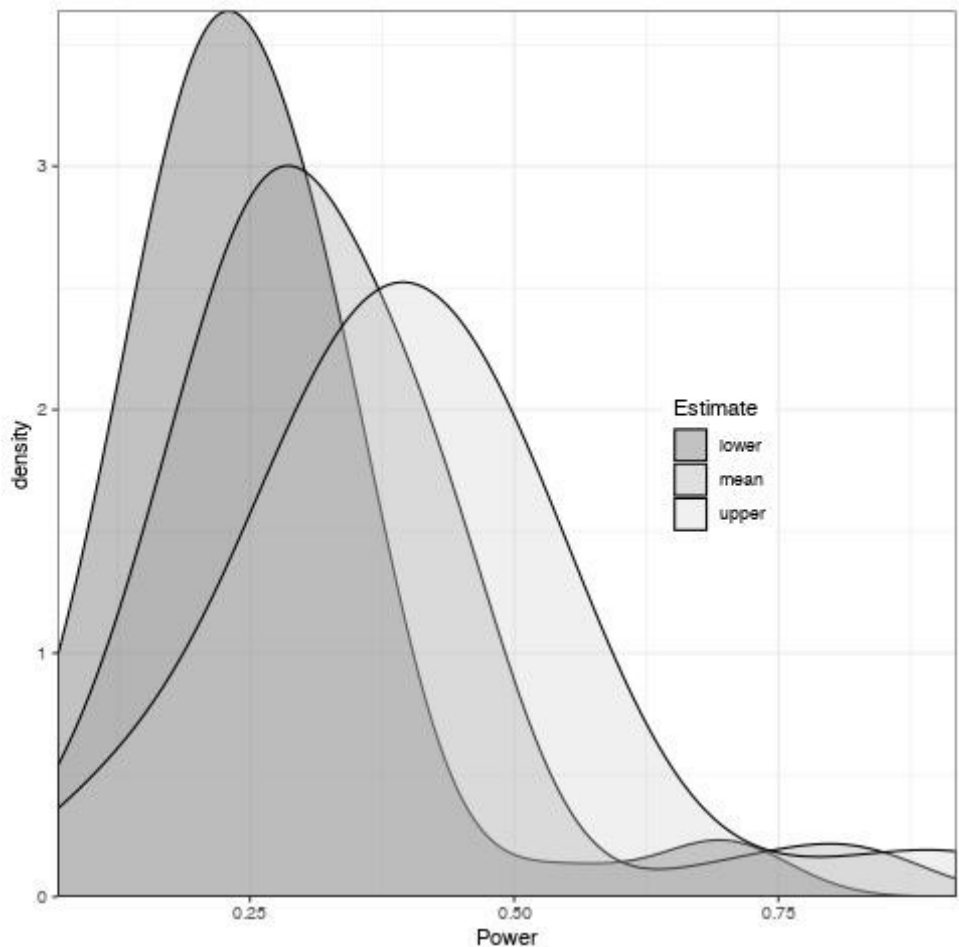

*Figure 4. Distribution of power across the 57 studies with raw datasets, considering an effect size of 21 ms (meta-analytic estimate obtained in Analysis 1) of 18 ms, and 24 ms (lower and upper bounds of the 95% credible interval, respectively).*

### *Analysis 3. Meta-regression with SOA as predictor*

The purpose of this analysis was to test whether the semantic interference differs at varying SOAs and justify the decision to then drop long SOAs from further analyses. To this end, SOA was transformed into a categorical variable. Amongst the 198 studies included in the meta-analysis (considering each SOA as a separate study), there were 99 datasets with an SOA of 0, 25 with short positive SOAs (from 43 to 150), 38 with short negative SOAs (from -43 to -160), 13 with long positive SOAs (from 200 to 1000 ms), and 23 with long negative SOAs (from -200 to -1000 ms). SOA was coded using treatment contrast such that each SOA is compared to an SOA of zero. The results of this analysis are presented in Table 3 and illustrated in Figure 5. A sensitivity analysis is presented in Appendix 4.

*Table 3. Results of the meta-regression testing for an effect of SOA on the semantic interference effect. The intercept corresponds to the size of the semantic interference effect at an SOA of 0, the other estimates indicate deviations from this estimate.*

|  | *Estimate (ms)* | *CrI (95%)* |
| --- | --- | --- |
| Intercept (SOA= zero) | 25.3 | 21.8 – 28.9 |
| Short negative SOA | -7.0 | -14.0 – -0.5 |
| Short positive SOA | -14.6 | -21.6 – -7.3 |
| Long negative SOA | -27.2 | -34.7 – -19.8 |
| Long positive SOA | -24.5 | -33.6 – -15.0 |
| Between-study SD | 13.0 | 10.9 – 15.5 |
| Number of studies | 198 | |

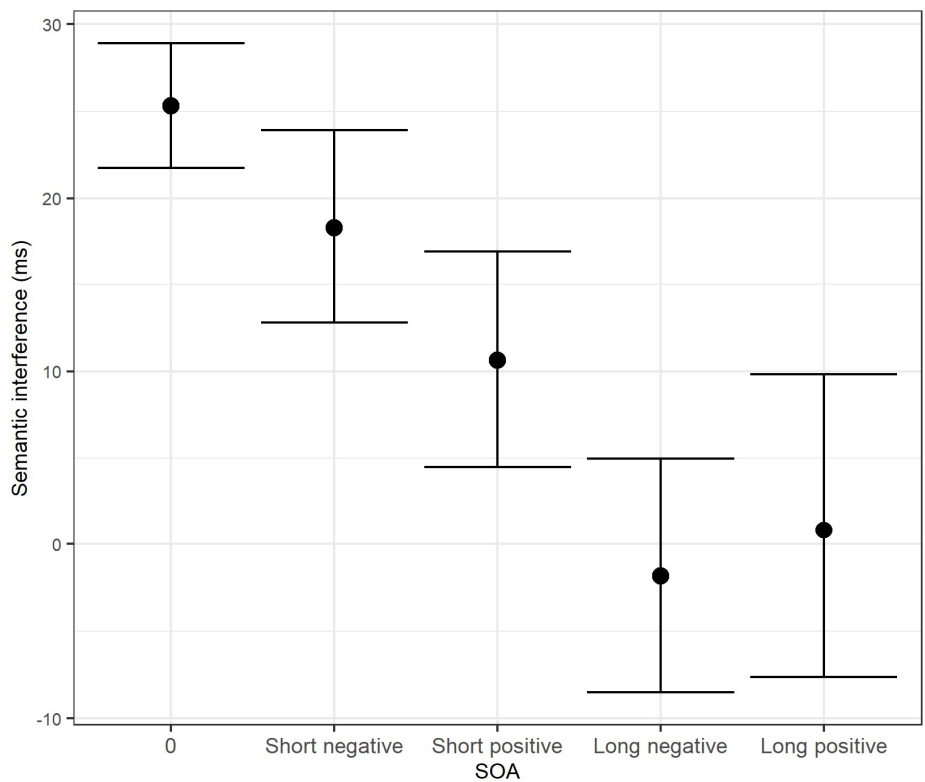

*Figure 5. Marginal effect, meta-regression testing the effect of SOA on the semantic interference effect*

The analysis reveals that the interference effect at an SOA of 0 is about 25 ms. This outcome provides support for the hypothesis that the semantic interference effect is found with SOAs of 0, short positive and short negative SOAs. At short positive SOAs the estimate of the semantic interference effect is more than half the size of the semantic interference effect at an SOA of zero. This analysis does not provide support for the hypotheses that semantic interference effects arise at long positive or negative SOAs or that the effect becomes facilitatory at such SOAs. In all subsequent analyses, the dataset was restricted to studies using short positive, short negative, and zero SOAs.

***Analysis 4. Meta-regression with properties of the design as predictors***

The next model included once again all studies (N = 162) with short SOAs (-160 to +150). We performed a meta-regression to explore the effects of the following predictors on the semantic interference effect: SOA, mean naming times in the unrelated condition, the interaction between SOA and mean naming times in the unrelated condition, familiarization, whether the distractors are part of the response set or not, set size (number of target words), number of repetitions of the target word set in the experiment, and number of semantic categories in the target set. In this analysis, SOA was entered as a continuous variable. *Number of repetitions of the target* word accounts for the number of times the participants named the pictures using the target set. Trials where participants produced the target words in a reading task (e.g., Glaser & Düngelhoff, 1984) or produced the target words in response to a different picture (Collina et al., 2013) were not taken into account. If the number of repetitions of the target set was mentioned by the authors, we used this value. When this information was not provided, we were able to infer it from the description of the design and procedure. The number of repetitions ranged from 1 to 48. *Familiarization* accounts for whether the participants were familiarized with the target words prior to the picture naming task. The value "yes" was given for all studies where a familiarization phase was mentioned, the value "no" for all others. We assume here that if a familiarization phase has been used, it is mentioned in the description of the procedure. There were 143 studies with a familiarization and 18 without (the information was missing for one unpublished datasets). *Distractors in response set* codes for whether the distractor and targets were the same or different words. In order to code for this variable, we first relied on the information provided by the authors. When this information was not provided we inferred it from the list of targets and distractors. There were 31 studies where distractors were part of the response set and 131 studies where different words were used for the target and distractor lists. The variable *Set size* corresponds to the number of different targets used in the experiment. *Mean Naming times in the unrelated condition* was either computed or extracted from the papers. It ranges from 570 ms to 1344 ms.

In order to obtain information about the *number of semantic categories in the target set* we first translated all items to English and generated semantic categories from the list of all English translations[4]. The resulting list of categories is presented in Appendix 5. Note that the definition of a semantic category is not straightforward. For instance, some authors consider that fruits and vegetables are two distinct categories, others that they both belong to the category FOOD. Moreover, where some authors see a category membership between two words, others might qualify the relationship as associative (e.g., *hairbrush* and *gel*) or as unrelated (e.g., *cave-valley*). As a consequence, our categories do not always match that used by the authors. Eighteen datasets had a missing value for this variable, as the authors did not report this information and did not provide the list of distractors.

---

[4] Note that our translations may sometimes diverge from that of the authors.

The main analysis included the 138 studies for which we had a value for all predictors. All continuous variables were centered around the mean. We used effect coding for familiarization and whether the distractors were part of the response set, with -0.5 for yes and 0.5 for no. As a consequence, the intercept in the model is the semantic interference effect when all variables are at their means. The distributions of the predictor values across studies is displayed in Figure 6. Table 4 displays the two-by-two correlations between these variables and the results of the statistical model are presented in Table 5. A sensitivity analysis for this model can be found in Appendix 6. Note that the same model without number of semantic categories (i.e., with the 24 additional studies for which this factor does not have any value) shows exactly the same results.

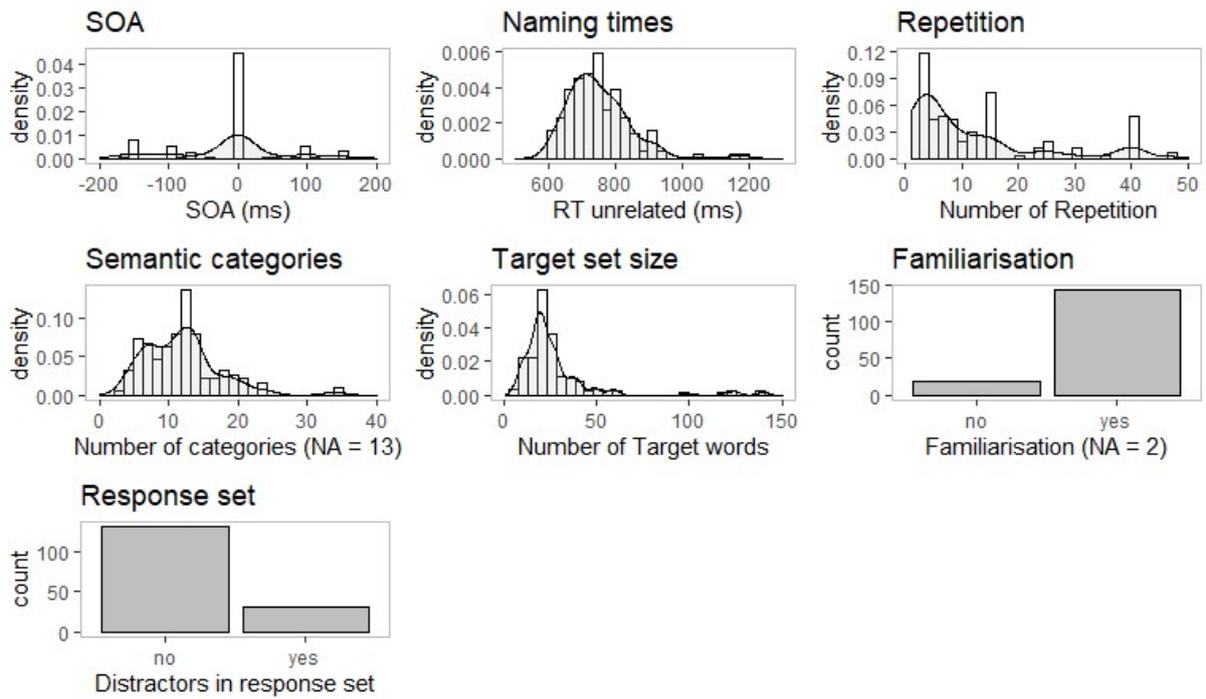

*Figure 6. Distribution of variables included as predictors in Analysis 4*

*Table 4. Pearson correlations between each pair of variables included in Analysis 4.*

|  | Set size | Naming times unrelated | Response set | SOA | Familiarization | Number of repetitions | Number of categories |
|---|---|---|---|---|---|---|---|
| Set size |  |  |  |  |  |  |  |
| Naming times unrelated | 0.34 |  |  |  |  |  |  |
| Response set | 0.01 | 0.16 |  |  |  |  |  |
| SOA | -0.12 | -0.13 | -0.05 |  |  |  |  |
| Familiarization | 0.08 | -0.06 | 0.06 | 0.02 |  |  |  |

| | | | | | | |
|---|---|---|---|---|---|---|
| Number of repetitions | -0.14 | -0.35 | 0.37 | 0.03 | 0.17 | |
| Number of categories | 0.44 | 0.26 | -0.38 | -0.09 | 0.1 | -0.32 |

*Table 5. Results of the meta-regression testing the role of aspects of the design on the semantic interference effect (analysis 4)*

| | Estimate (ms) | CrI (95%) |
|---|---|---|
| Intercept | 20.0 | 12.9 – 26.9 |
| SOA (centered) | -0.03 | -0.08 – 0.02 |
| Naming times unrelated (centered) | 0.09 | 0.04 – 0.1 |
| Familiarization | -2.7 | -15.6 – 10.5 |
| Number of categories (centered) | 0.08 | -0.6 – 0.7 |
| Number of repetitions (centered) | -0.19 | -0.5 – 0.1 |
| Set size (centered) | -0.05 | -0.3 – 0.2 |
| Response set | -2.0 | -10.5 – 6.9 |
| SOA * RT unrelated | -0.0003 | -0.0008 – 0.0003 |
| Number of studies | 138 | |
| Between study SD | 11.5 | 9.2 – 14.2 |

As can be seen in Table 5, the 95% credible intervals for all estimates except for the predictor *Naming times unrelated* include zero. The overall semantic interference effect is similar in size and precision to that of the model without predictors (Analysis 1), similarly, the between-study standard deviations in the two models are about the same size. The same model was run without the predictor *Naming times unrelated* to determine whether the effect of this predictor was a by-product of its correlation with other predictors. In the model without this predictor, the credible interval of the estimate of the predictor *Number of repetition of the target word* ranges from -0.7 to -0.12 ms. According to this model, the semantic interference effect decreases by 0.40 ms for each additional repetition.

***Analysis 5. Meta-analyses of interactions with distractor frequency, distractor length, and distractor picturability***

The purpose of this series of meta-analyses was to examine the role of the properties of the distractors on the semantic interference effect. We focused on distractor length (i.e., number of letters), distractor frequency, and picturability[5]. We restricted the dataset to the studies for which we had the raw data. We had to exclude the studies by Starreveld (2013) and Shao (2015) given that we did not have information about the distractors used in individual trials. Moreover, the studies by Zhang et al. (2016) were not included in the model with distractor length, as all distractors had the same length (two Chinese characters) in these experiments. Finally, in the meta-analysis involving picturability, we only considered the datasets whose distractors exhibited a certain amount of variability in the picturability ratings.

---

[5] Miozzo and Caramazza (2003) reported a decrease of the semantic interference effect with the repetition of the target-distractor stimuli. We conducted a meta-analysis of the interaction between this variable and semantic interference using the seven raw datasets where target-word stimuli were repeated. The results are presented in the supplementary material provided with this paper.

We extracted the lexical frequency of the distractors from the following corpora. The Subtlex corpora of subtitle frequencies were used for German (190,500 words), Dutch (437,503 words), Italian (517,564 words), Spanish (94,338 words), Chinese Mandarin (99,121 words), US English (74,286 words), and UK English (160,022 words). Subtitle frequencies were used (see Brysbaert, Buchmeier, et al., 2011; Brysbaert, Keuleers, & New, 2011; Cai & Brysbaert, 2010; Cuetos, GlezNosti, Barbón, & Brysbaert, 2011; New, Brysbaert, Veronis, & Pallier, 2007 for evidence suggesting that the best predictors of lexical decision and word naming are those computed on large corpora of subtitles). For French, we used lexeme subtitle frequency from the database Lexique (142,694 words, New, Pallier, Brysbaert, & Ferrand, 2004). The number of letters was taken as a measure of word length.

To obtain some measure of picturability, the first three authors as well as three additional individuals rated the picturability of all distractors on a 6-point scale. Following Paivio (1971) the ratings reflected the ease and speed of the mental image associated with each written word. Distractor words that quickly generated a clear mental image were given a score of 6, this number decreased if the mental image was less clear or took more time to form. Distractors for which no mental image could be generated were given a 1. We then computed, for each word, the mean of all ratings and disregarded the five studies with no distractor rated below 4.5. The distributions of the mean frequency, mean number of letters, and mean picturability across studies are displayed in Figure 7.

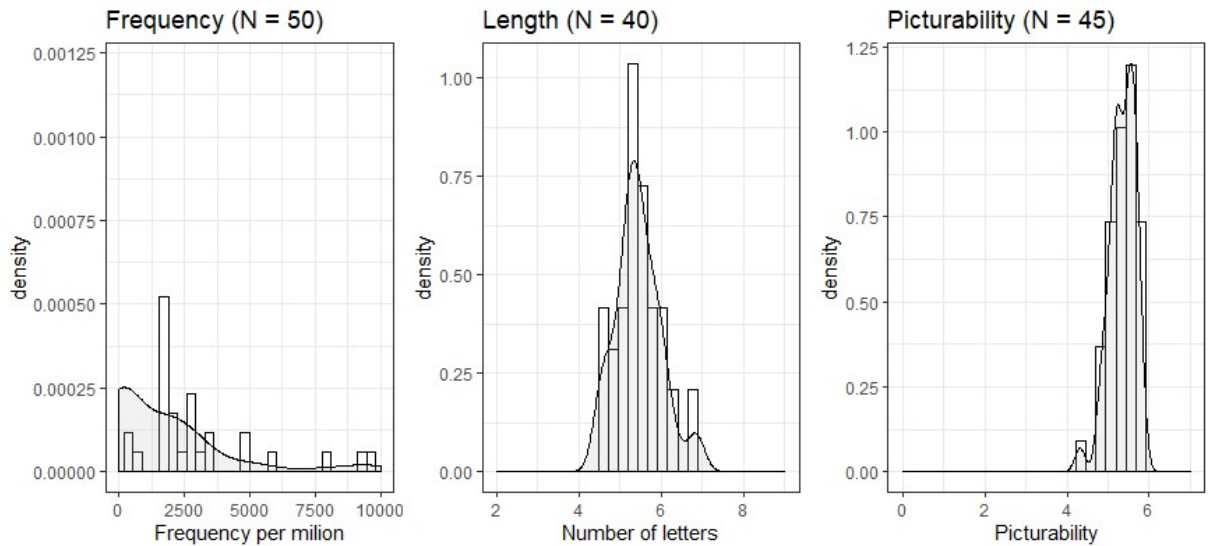

*Figure 7. Distribution of mean number of letters, frequency and picturability across datasets*

The correlation between mean lexical frequency and mean number of letters was equal to -0.18, that between mean lexical frequency and mean picturability to 0.20 and that between mean number of letters and picturability to -0.15.

For each study, we conducted three mixed-effects model (using lme4). In the first we included the contrast between semantically related and unrelated distractors, distractor frequency (log of frequency per million +1), and their interaction, in the second the contrast between semantically related and unrelated distractors, distractor length (centered around the mean) and their interaction, and in the third the contrast between semantically related and unrelated distractors, distractor picturability (centered around the mean) and their interaction. We then examined the estimates of each interaction in separate Bayesian meta-analyses. In these analyses, a positive intercept for the interaction suggests that an increase in the values of the predictor results in an increase of the semantic interference effect. Table 6 summarizes the results of these meta-analyses and

Figure 8 displays the posterior distributions of the estimates. Figure 9 displays, for each study, the estimate of the interaction between semantic interference and distractor frequency in the data as well as weighted by the meta-analysis. Similar figures for the interactions involving distractor length and distractor picturability can be found in the supplementary material. Sensitivity analyses were conducted for each meta-analysis. The results of these analyses are detailed in **Appendix 7**.

*Table 6. Output of the meta-analyses of the interactions between semantic relatedness and distractor frequency, distractor length, and distractor picturability*

|  |  | *Estimate (ms)* | *CrI (95%)* |
|---|---|---|---|
| Distractor frequency | Intercept | - 0.8 | -3.3 – 1.7 |
|  | Between-study SD | 2.2 | 0.09 – 6.0 |
|  | Number of studies | 50 |  |
| Distractor length | Intercept | 2.7 | -0.1 – 5.5 |
|  | Between-study SD | 2.5 | 0.1 – 6.9 |
|  | Number of studies | 40 |  |
| Distractor picturability | Intercept | -0.3 | -7.8 – 5.9 |
|  | Between-study SD | 10.5 | 0.9 – 21.5 |
|  | Number of studies | 45 |  |

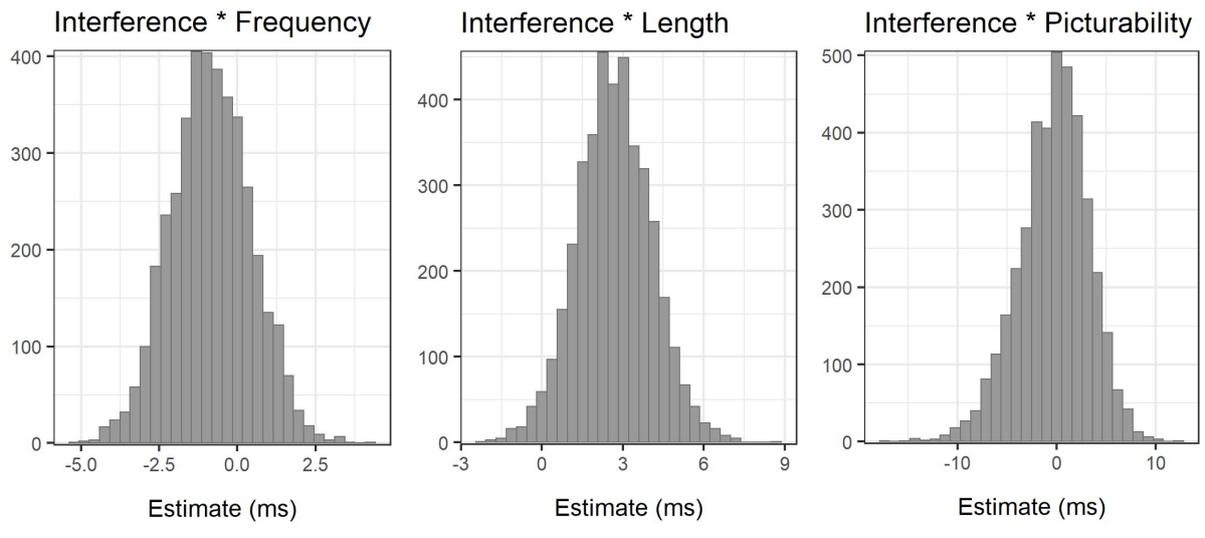

*Figure 8. Posterior distributions of the interactions between distractor frequency, distractor length, distractor picturability, and semantic relatedness.*

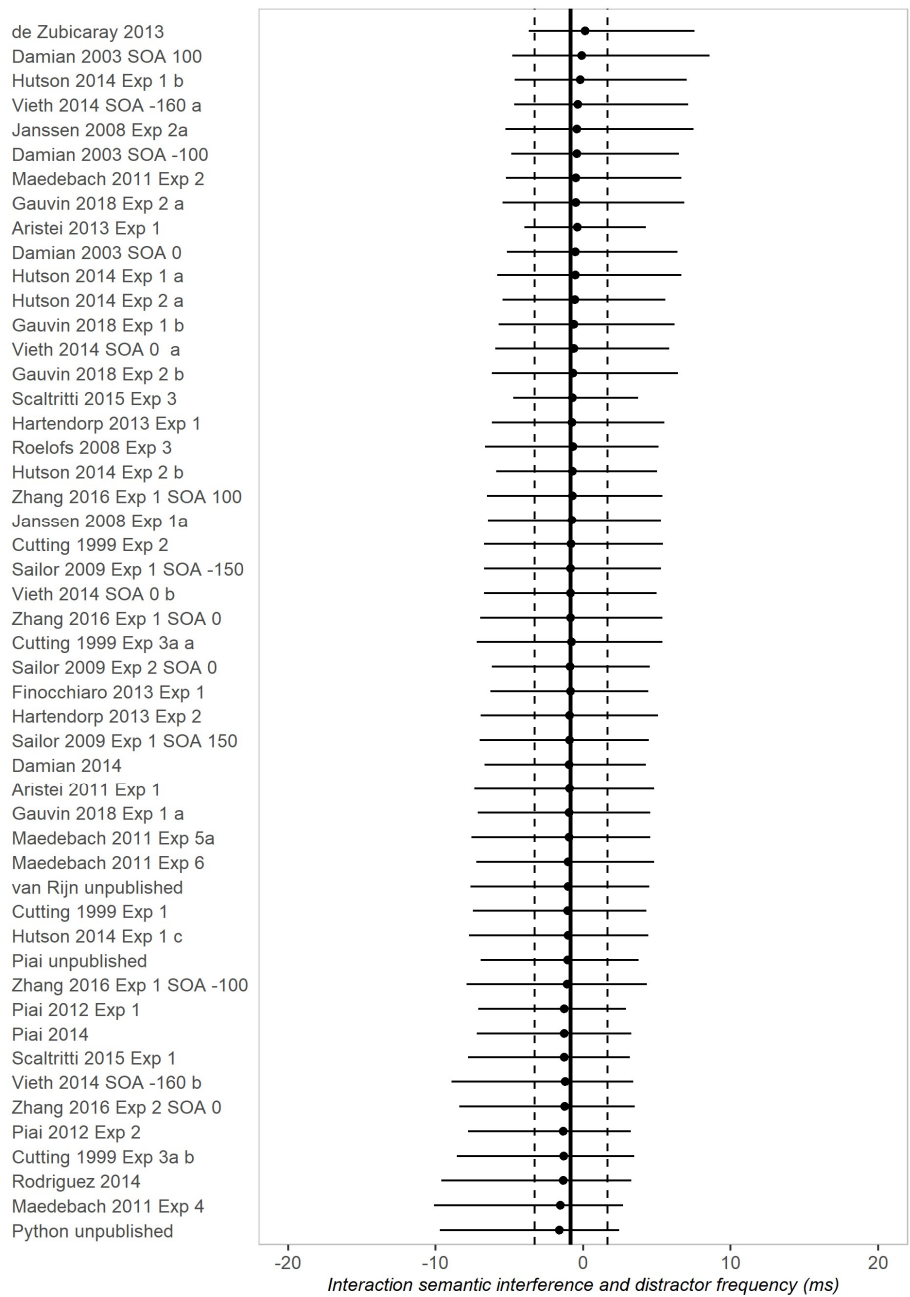

*Figure 9. Summary of the random-effects meta-analysis modeling the interaction between distractor frequency and semantic relatedness. For each experiment, the figure displays the mean and posterior estimate (mean and 95% credible interval). A positive value means that distractors with higher frequency show a stronger interference effect. The black vertical line represents the grand mean (i.e., the meta-analytic effect) and the dashed vertical lines delimit the 95% credible interval of that estimate. Details on the different studies are to be found in the Supplementary material.*

To summarize, the 95% credible intervals for the estimate of the interaction between distractor frequency and semantic relatedness ranges between -3 and 2 ms, that of the interaction between distractor length and semantic relatedness between -0.1 and 5 ms and that of the interaction between distractor picturability and semantic relatedness between -8 to 6 ms.

**General discussion**

A first aim of this contribution was to provide an estimate of the size of the semantic interference effect together with information about its degree of uncertainty, considering all the available evidence to date. The posterior distribution of the meta-analysis including 162 studies with SOAs of 0 or short SOAs reveals that the effect is about 21 ms with a 95% credible interval ranging from 18 to 24 ms. This result confirms that overall and taking into account studies with different designs, languages, and materials, distractors of the same semantic category interfere more with the preparation of the response than distractors of a different semantic category. This meta-analysis further reveals substantial variability across studies, suggesting that studies differ in the underlying true effect, as a likely consequence of the many differences in the design or materials used across studies (many studies involved for instance additional relationships between target words and distractors, e.g., semantic associates, phonological overlap etc.).

A second aim of this contribution was to provide information on the power of previous studies. Our analysis shows that the majority of the studies testing for a semantic interference effect in our database were likely underpowered. This is problematic because it makes it difficult to interpret the results of these studies, especially the null findings. In the light of this information, we can speculate that non-significant effects in a subset of studies or conditions (e.g., with specific SOAs, specific set sizes, etc.) are type II errors.

The third aim of this study was specifically to explore claims in the literature about the variables that modulate the semantic interference effect, when considering all the relevant evidence published to date. In the present discussion, and for the sake of simplicity, we will consider that estimates whose 95 % credible intervals do not cross zero provide support for the hypothesis that the corresponding variables modulate the semantic interference effect. Note however that this criterion is merely a working tool and readers are encouraged to adopt their own criteria.

A first meta-regression confirms the role of SOA and the claim that the semantic interference effect is observed when the picture and distractor are presented at the same time or within 100-160 ms from one another. The outcome of this analysis further suggests that the semantic interference effect is greater when the distractor word and the picture are displayed at the same time than when the distractor is presented shortly after the picture. At long negative SOAs, the estimate of the semantic facilitation is negative but has a size of 2 ms, with a credible interval between -9 and 5 ms. Our analysis thus does not allow us to confirm (or refute) the claim that long negative SOAs produce facilitation. Note however that the number of datasets in which such long SOAs were used is rather small.

We further explored the impact of several properties of the design or of the materials on the semantic interference effect. We first investigated the effects of familiarization, number of target words, number of semantic categories, number of repetitions, mean naming times in the unrelated condition, the interaction between the latter and SOA, and whether the distractor was part of the response set. The results of the meta-regression suggest that the semantic interference effect increases with longer naming times in the unrelated condition but does not allow us to confirm (or refute) the hypothesis that naming times interact with SOA. If we remove naming times from the model, the number of repetitions of the target word is found to modulate the interference effect, with a decrease in semantic interference effect with each subsequent production of the target word. This suggests that naming times is only a relevant factor via its correlation with repetition. The modulation of the semantic interference effect by repetition can be explained by any account of the semantic interference effect where this effect can be modulated by the relative speed of processing/encoding between the distractor and target words. The time needed to encode a word decreases with repetition (e.g., Belke et al., 2005; Bürki & Laganaro, 2014). As a consequence, repetition increases the delay between the encoding of the target word and the processing of the distractor word. Under a lexical-competition account, the semantic information related to the distractor might become active too late to interfere with the encoding of the target word. In a response-exclusion account, the repeated target word will enter the response buffer earlier, the probability that the distractor can enter the response buffer before the target word will therefore be lower for trials with repeated target words.

For all other variables, the meta-regression does not allow us to confirm (or refute) the hypothesis that they modulate the semantic interference effect. Importantly, however, some of these variables (including familiarization, or whether the distractors are part of the response set) showed little variation across datasets and clearly require additional empirical evidence. Regarding the role of familiarization, it must further be noted that many of the studies without a familiarization phase presented the target-distractors pairs multiple times (e.g., at different SOAs, e.g., Dell'Acqua et al., 2007; with clearly visible and masked distractors in Finkbeiner and Caramazza, 2006). It could be argued that the first repetition of the target words has the same effect as a familiarization phase.

More generally, the results of any meta-regression should be considered exploratory, especially when the predictors were not directly manipulated in the studies themselves. Finally, it could be that our model was too simplistic and that the effects of these variables can only be found when considering their interactions with one another. According to Roelofs' model of the semantic interference effect for instance, these variables are involved in complex interactions. We did not investigate these interactions, the data were too scarce and the predictions too broad. This said, the output of our analysis is in line with the qualitative review, where the statistical evidence in favor of interactions between each of these variables and semantic relatedness is either lacking or ambiguous. Taken together, the outcome of the meta-regression and the qualitative review force us to conclude that the available evidence to date is not sufficient to conclude that familiarization, the number of target words, the number of semantic categories, or whether the distractor is part of the response set, truly modulate the semantic interference effect.

We then examined the impact of the properties of the distractors on the semantic interference effect. Note that unlike in the meta-regression mentioned above, these were examined in a series of meta-analyses, each based on the information provided by between 40 and 50 raw datasets. At the end of the Introduction, we discussed possible predictions of lexical-competition and response exclusion accounts regarding interactions between the properties of the distractors and the semantic interference effect. In both accounts, a weaker semantic interference effect can be expected for distractors that are processed more slowly. The meta-analyses of the interactions between semantic relatedness and frequency, length, or picturability do not provide support for or against the hypotheses that these variables modulate the semantic interference effect.

To summarize, our analyses suggest that the semantic interference effect is modulated by SOA and the number of repetitions of the target word. It remains an open issue whether other variables also modulate the semantic interference effect. A modulation of the semantic interference effect by SOA and repetition is predicted or can be explained by all accounts.

Taken together, the qualitative review and series of meta-analyses and meta-regressions clearly suggest that after 40 years of research, the empirical evidence regarding the factors that modulate the semantic interference effect is not sufficient to decide whether the semantic interference effect arises during lexical selection or later. The qualitative review further highlights the often difficult endeavor of mapping theoretical accounts to specific predictions. If the factors that modulate the semantic interference effect can truly be used to determine the locus of this effect, this endeavor will only be successful if specific predictions for all these variables are derived from each account a priori, and tested in sufficiently powered and carefully controlled experiments.

Two theoretically important variables were not included in the present meta-analysis and should be given some attention in future work. Several studies have argued that the semantic interference effect should increase with semantic distance in lexical-competition models, but should not modulate the interference effect according to the response-exclusion account. The literature offers a mixed picture, with some studies showing more interference for close pairs, other studies showing less interference for these same pairs, and failed attempts to replicate these findings. More research is needed to clarify the role of this variable as well as to determine how to implement the notion of semantic distance such that it represents the organization of conceptual/semantic information in the speaker's mind. The second variable that deserves attention is the number of items in the semantic categories. This variable can be expected to modulate the semantic interference effect in the swinging lexical network of Abdel Rahman and Melinger (e.g., Abdel Rahman & Melinger, 2019). In this model, the amount of interference depends on the number of active competitors. Related distractors of categories with many members are expected to activate a larger network of interrelated competitors than distractors of categories with fewer members. There are no a priori reasons to assume that this factor should modulate the semantic interference effect in the response-exclusion hypothesis or in other versions of the competition account. To the best of our knowledge, the impact of this variable has not yet been investigated and would offer an interesting test case for the swinging lexical network.

**Conclusion**

Taking over 150 studies on the semantic interference effect in the picture-word interference paradigm into account, we provide an estimate of this effect. We explore the factors that modulate the semantic interference effect and show that despite the high number of studies devoted to this issue in the last 40 years, no definitive conclusion is warranted. The available evidence to date offers little information about the nature of the semantic interference effect and can hardly be used to model the nature of lexical access.

**References**

Abdel Rahman, R., & Aristei, S. (2010). Now you see it … and now again: Semantic interference reflects lexical competition in speech production with and without articulation. Psychonomic Bulletin & Review, 17(5), 657–661. https://doi.org/10.3758/PBR.17.5.657

Abdel Rahman, R., & Melinger, A. (2007). When bees hamper the production of honey: Lexical interference from associates in speech production. Journal of Experimental Psychology. Learning, Memory, and Cognition, 33(3), 604–614. https://doi.org/10.1037/0278-7393.33.3.604

Abdel Rahman, R., & Melinger, A. (2009). Semantic context effects in language production: A swinging lexical network proposal and a review. Language and Cognitive Processes, 24(5), 713–734. https://doi.org/10.1080/01690960802597250

Abdel Rahman, R., & Melinger, A. (2019). Semantic processing during language production: An update of the swinging lexical network. Language, Cognition and Neuroscience, 34(9), 1176-1192. https://doi.org/10.1080/23273798.2019.1599970

Abel, S., Dressel, K., Bitzer, R., Kümmerer, D., Mader, I., Weiller, C., & Huber, W. (2009). The separation of processing stages in a lexical interference fMRI-paradigm. NeuroImage, 44, 1113-1124. https://doi.org/10.1016/j.neuroimage.2008.10.018

Alario, F. X., Segui, J., & Ferrand, L. (2000). Semantic and associative priming in picture naming. The Quarterly Journal of Experimental Psychology: Human Experimental Psychology, 53(3), 741–764. https://doi.org/10.1080/713755907

Aristei, S., Melinger, A., & Abdel Rahman, R. (2011). Electrophysiological chronometry of semantic context effects in language production. Journal of Cognitive Neuroscience, 23(7), 1567–1586. https://doi.org/10.1162/jocn.2010.21474

Aristei, S., & Abdel Rahman, R. (2013). Semantic interference in language production is due to graded similarity, not response relevance. Acta Psychologica, 144(3), 571–582. https://doi.org/10.1016/j.actpsy.2013.09.006

Barton, J. J. S., Hanif, H. M., Björnström, L. E., & Hills, C. (2014). The word-length effect in reading: A review. Cognitive neuropsychology, 31(5–6), 378–412. https://doi.org/10.1080/02643294.2014.895314

Bates, D., Mächler, M., Bolker, B., & Walker, S. (2015). Fitting linear mixed-effects models using lme4. Journal of Statistical Software, 67(1), 1–48. https://doi.org/10.18637/jss.v067.i01

Belke, E., Brysbaert, M., Meyer, A. S., & Ghyselinck, M. (2005). Age of acquisition effects in picture naming: Evidence for a lexical-semantic competition hypothesis. Cognition, 96(2), B45–B54. https://doi.org/10.1016/j.cognition.2004.11.006

Bi, Y., Xu, Y., & Caramazza, A. (2009). Orthographic and phonological effects in the picture–word interference paradigm: Evidence from a logographic language. Applied Psycholinguistics, 30(04), 637–658. https://doi.org/10.1017/S0142716409990051

Blackford, T., Holcomb, P. J., Grainger, J., & Kuperberg, G. R. (2012). A funny thing happened on the way to articulation: N400 attenuation despite behavioral interference in picture naming. Cognition, 123(1), 84–99. https://doi.org/10.1016/j.cognition.2011.12.007

Bloem, I., & La Heij, W. (2003). Semantic facilitation and semantic interference in word translation: Implications for models of lexical access in language production. Journal of Memory and Language, 48(3), 468–488. https://doi.org/10.1016/S0749-596X(02)00503-X

Brysbaert, M., Buchmeier, M., Conrad, M., Jacobs, A. M., Bölte, J., & Böhl, A. (2011). The word frequency effect: A review of recent developments and implications for the choice of frequency estimates in German. Experimental Psychology, 58(5), 412–424. https://doi.org/10.1027/1618-3169/a000123

Brysbaert, M., Keuleers, E., & New, B. (2011). Assessing the usefulness of Google books' word frequencies for psycholinguistic research on word processing. Frontiers in Psychology, 2. https://doi.org/10.3389/fpsyg.2011.00027

Brysbaert, M., & Stevens, M. (2018). Power analysis and effect size in mixed effects models: A Tutorial. Journal of Cognition, 1(1). https://doi.org/10.5334/joc.10

Bürki, A., & Laganaro, M. (2014). Tracking the time course of multi-word noun phrase production with ERPs or on when (and why) cat is faster than the big cat. Frontiers in Psychology, 5, 586. https://doi.org/10.3389/fpsyg.2014.00586

Bürkner, P. (2018). Advanced bayesian multilevel modeling with the R Package brms. *The R Journal*, 10(1), 395–411. https://doi.org/10.32614/RJ-2018-017.

Cai, Q., & Brysbaert, M. (2010). SUBTLEX-CH: Chinese word and character frequencies based on film subtitles. PloS One, *5*(6), e10729. https://doi.org/10.1371/journal.pone.0010729

Caramazza, A. (1997). How many levels of processing are there in lexical access? Cognitive Neuropsychology, 14(1), 177–208. https://doi.org/10.1080/026432997381664

Caramazza, A., & Costa, A. (2000). The semantic interference effect in the picture-word interference paradigm: Does the response set matter? Cognition, 75(2), B51–B64. https://doi.org/10.1016/S0010-0277(99)00082-7

Caramazza, A., & Costa, A. (2001). Set size and repetition in the picture--word interference paradigm: Implications for models of naming. Cognition, 80(3), 291–298. https://doi.org/10.1016/s0010-0277(00)00137-2

Collina, S., Tabossi, P., & Simone, F. D. (2013). Word production and the picture-word interference paradigm: The role of learning. Journal of Psycholinguistic Research, 42(5), 461–473. https://doi.org/10.1007/s10936-012-9229-z

Collins, A. M., & Loftus, E. F. (1975). A spreading-activation theory of semantic processing. Psychological Review, 82, 407-428. https://doi.org/10.1037/0033-295X.82.6.407

Costa, A., Alario, F.-X., & Caramazza, A. (2005). On the categorical nature of the semantic interference effect in the picture-word interference paradigm. Psychonomic Bulletin & Review, 12(1), 125–131. https://doi.org/10.3758/BF03196357

Costa, A., & Caramazza, A. (1999). Is lexical selection in bilingual speech production language-specific? Further evidence from Spanish–English and English–Spanish bilinguals. Bilingualism: Language and Cognition, 2(3), 231–244. https://doi.org/10.1017/S1366728999000334

Cuetos, F., Glez-Nosti, M., Barbón, A., & Brysbaert, M. (2011). SUBTLEX-ESP: Spanish word frequencies based on film subtitles. Psicológica, 32(2), 133–143.

Cutting, J. C., & Ferreira, V. S. (1999). Semantic and phonological information flow in the production lexicon. Journal of Experimental Psychology. Learning, Memory, and Cognition, 25(2), 318–344. https://doi.org/10.1037//0278-7393.25.2.318

Damian, M. F., & Bowers, J. S. (2003). Locus of semantic interference in picture-word interference tasks. Psychonomic Bulletin & Review, 10(1), 111–117. https://doi.org/10.3758/BF03196474

Damian, M. F., & Martin, R. C. (1999). Semantic and phonological codes interact in single word production. Journal of Experimental Psychology. Learning, Memory, and Cognition, 25(2), 345–361. https://doi.org/10.1037//0278-7393.25.2.345

Damian, M. F., & Spalek, K. (2014). Processing different kinds of semantic relations in picture-word interference with non-masked and masked distractors. Frontiers in Psychology, 5. https://doi.org/10.3389/fpsyg.2014.01183

Dell, G. S. (1986). A spreading-activation theory of retrieval in sentence production. Psychological Review, 93(3), 283–321.

Dell'Acqua, R., Job, R., Peressotti, F., & Pascali, A. (2007). The picture-word interference effect is not a Stroop effect. Psychonomic Bulletin & Review, 14(4), 717–722. https://doi.org/10.3758/bf03196827

Dhooge, E., & Hartsuiker, R. J. (2010). The distractor frequency effect in picture-word interference: Evidence for response exclusion. Journal of Experimental Psychology. Learning, Memory, and Cognition, 36(4), 878–891. https://doi.org/10.1037/a0019128

Diaz, M. T., Hogstrom, L. J., Zhuang, J., Voyvodic, J., Johnson, M. A., & Camblin, C. C. (2014). Written distractor words influence brain activity during overt picture naming. Frontiers in Human Neurosciences, 44, 1113–1124. https://doi.org/10.3389/fnhum.2014.00167

Finkbeiner, M., & Caramazza, A. (2006). Now you see it, now you don't: On turning semantic interference into facilitation in a stroop-like task. Cortex, 42(6), 790–796. https://doi.org/10.1016/S0010-9452(08)70419-2

Finkbeiner, M., Forster, K., Nicol, J., & Nakamura, K. (2004). The role of polysemy in masked semantic and translation priming. Journal of Memory and Language, 51(1), 1–22. https://doi.org/10.1016/j.jml.2004.01.004

Finocchiaro, C., & Navarrete, E. (2013). About the locus of the distractor frequency effect: Evidence from the production of clitic pronouns. Journal of Cognitive Psychology, 25(7), 861–872. https://doi.org/10.1080/20445911.2013.832254

Gauvin, H. S., Jonen, M. K., Choi, J., McMahon, K., & de Zubicaray, G. I. (2018). No lexical competition without priming: Evidence from the picture–word interference paradigm. Quarterly Journal of Experimental Psychology, 174702181774726. https://doi.org/10.1177/1747021817747266

Gelman, A., & Carlin, J. (2014). Beyond power calculations: Assessing Type S (Sign) and Type M (Magnitude) Errors. Perspectives on Psychological Science, 9(6), 641–651. https://doi.org/10.1177/1745691614551642

Glaser, W. R., & Düngelhoff, F.-J. (1984). The time course of picture-word interference. Journal of Experimental Psychology: Human Perception and Performance, 10(5), 640–654. https://doi.org/10.1037/0096-1523.10.5.640

Glaser, W. R., & Glaser, M. O. (1989). Context effects in Stroop-like word and picture processing. Journal of Experimental Psychology: General, 118(1), 13–42. https://doi.org/10.1037/0096-3445.118.1.13

Goldrick, M. (2006). Limited interaction in speech production: Chronometric, speech error, and neuropsychological evidence. Language and Cognitive Processes, 21(7–8), 817–855. https://doi.org/10.1080/01690960600824112

Hantsch, A., Jescheniak, J. D., & Schriefers, H. (2005). Semantic competition between hierarchically related words during speech planning. Memory & Cognition, 33(6), 984–1000. https://doi.org/10.3758/BF03193207

Hartendorp, M. O., Van der Stigchel, S., & Postma, A. (2013). To what extent do we process the nondominant object in a morphed figure? Evidence from a picture–word interference task. *Journal of Cognitive Psychology, 25*(7), 843–860. https://doi.org/10.1080/20445911.2013.832197

Hashimoto, N., & Thompson, C. K. (2010). The use of the picture–word interference paradigm to examine naming abilities in aphasic individuals. Aphasiology, 24(5), 580–611. https://doi.org/10.1080/02687030902777567

Hauk, O., Coutout, C., Holden, A., & Chen, Y. (2012). The time-course of single-word reading: Evidence from fast behavioral and brain responses. Neuroimage, 60(2–2), 1462–1477. https://doi.org/10.1016/j.neuroimage.2012.01.061

Hirschfeld, G., Jansma, B., Bölte, J., & Zwitserlood, P. (2008). Interference and facilitation in overt speech production investigated with event-related potentials. Neuroreport, 19(12), 1227–1230. https://doi.org/10.1097/WNR.0b013e328309ecd1

Hoenig, J. M., & Heisey, D. M. (2001) The Abuse of Power, The American Statistician, 55(1), 19-24, https://doi.org/10.1198/000313001300339897

Hutson, J., & Damian, M. F. (2014). Semantic gradients in picture-word interference tasks: Is the size of interference effects affected by the degree of semantic overlap? Frontiers in Psychology, 5, 872. https://doi.org/10.3389/fpsyg.2014.00872

Indefrey, P. (2011). The spatial and temporal signatures of word production components: A critical update. Frontiers in Psychology, 2. https://doi.org/10.3389/fpsyg.2011.00255

Janssen, N., Schirm, W., Mahon, B. Z., & Caramazza, A. (2008). Semantic Interference in a Delayed Naming Task: Evidence for the Response Exclusion Hypothesis. Journal of Experimental Psychology. Learning, memory, and cognition, 34(1), 249–256. https://doi.org/10.1037/0278-7393.34.1.249

Judd, C. M., Westfall, J., & Kenny, D. A. (2012). Treating stimuli as a random factor in social psychology: A new and comprehensive solution to a pervasive but largely ignored problem. Journal of Personality and Social Psychology, 103(1), 54–69. https://doi.org/10.1037/a0028347

Klein, G. S. (1964). Semantic power measured through the interference of words with color-naming. The American Journal of Psychology, 77(4), 576–588. https://doi.org/10.2307/1420768

Kleinman, D. (2013). Resolving semantic interference during word production requires central attention. Journal of experimental psychology. Learning, memory, and cognition, 39(6). https://doi.org/10.1037/a0033095

Krott, A., Medaglia, M. T., & Porcaro, C. (2019). Early and late effects of semantic distractors on electroencephalographic responses during overt picture naming. Frontiers in Psychology, 10. https://doi.org/10.3389/fpsyg.2019.00696

Kuipers, J.-R., La Heij, W. L., & Costa, A. (2006). A further look at semantic context effects in language production: The role of response congruency. Language and Cognitive Processes, 21(7–8), 892–919. https://doi.org/10.1080/016909600824211

La Heij, W. (1988). Components of Stroop-like interference in picture naming. Memory & Cognition, 16(5), 400–410. https://doi.org/10.3758/BF03214220

La Heij, W., Dirkx, J., & Kramer, P. (1990). Categorical interference and associative priming in picture naming. British Journal of Psychology, 81(4), 511–525. https://doi.org/10.1111/j.2044-8295.1990.tb02376.x

La Heij, W., & Hof, E. van den. (1995). Picture-word interference increases with target-set size. Psychological Research, 58(2), 119–133. https://doi.org/10.1007/BF00571100

La Heij, W., & Vermeij, M. (1987). Reading versus naming: The effect of target set size on contextual interference and facilitation. Perception & Psychophysics, 41(4), 355–366. https://doi.org/10.3758/BF03208237

Laganaro, M., & Alario, F.-X. (2006). On the locus of the syllable frequency effect in speech production. Journal of Memory and Language, 55(2), 178–196. https://doi.org/10.1016/j.jml.2006.05.001

Lamers, M. J. M., Roelofs, A., & Rabeling-Keus, I. M. (2010). Selective attention and response set in the Stroop task. Memory & Cognition, 38(7), 893–904. https://doi.org/10.3758/MC.38.7.893

Levelt, W. J. M., Roelofs, A., & Meyer, A. S. (1999). A theory of lexical access in speech production. The Behavioral and Brain Sciences, 22(1), 1–38; discussion 38-75. https://doi.org/10.1017/s0140525x99001776

Lupker, S. J. (1982). The role of phonetic and orthographic similarity in picture-word interference. Canadian Journal of Psychology, 36(3), 349–367. https://doi.org/10.1037/h0080652

Lupker, S. J. (1979). The semantic nature of response competition in the picture-word interference task. Memory & Cognition, 7(6), 485–495. https://doi.org/10.3758/BF03198265

Mädebach, A., Oppermann, F., Hantsch, A., Curda, C., & Jescheniak, J. D. (2011). Is there semantic interference in delayed naming? Journal of Experimental Psychology. Learning, Memory, and Cognition, 37(2), 522–538. https://doi.org/10.1037/a0021970

McRae, K., de Sa, V. R., & Seidenberg, M. S. (1997). On the nature and scope of featural representations of word meaning. Journal of Experimental Psychology: General, *126*(2), 99–130. https://doi.org/10.1037/0096-3445.126.2.99

Mahon, B. Z., Costa, A., Peterson, R., Vargas, K. A., & Caramazza, A. (2007). Lexical selection is not by competition: A reinterpretation of semantic interference and facilitation effects in the picture-word interference paradigm. Journal of Experimental Psychology. Learning, Memory, and Cognition, 33(3), 503–535. https://doi.org/10.1037/0278-7393.33.3.503

Meyer, A. S. (1996). Lexical access in phrase and sentence production: Results from picture–word interference experiments. Journal of Memory and Language, 35(4), 477–496. https://doi.org/10.1006/jmla.1996.0026

Miozzo, M., & Caramazza, A. (2003). When more is less: A counterintuitive effect of distractor frequency in the picture-word interference paradigm. Journal of Experimental Psychology. General, 132(2), 228–252. https://doi.org/10.1037/0096-3445.132.2.228

Morey, R. D., Hoekstra, R., Rouder, J. N., Lee, M.D., & Wagenmakers, E. –J. (2016). The fallacy of placing confidence in confidence intervals. Psychonomic Bulletin & Review, 23, 103-123. https://doi.org/10.3758/s13423-015-0947-8

Open Science Collaboration (2015). Estimating the reproducibility of psychological science. Science*, 349(6251), aac4716. https://doi.org/10.1126/science.aac4716

New, B., Brysbaert, M., Veronis, J., & Pallier, C. (2007). The use of film subtitles to estimate word frequencies. Applied Psycholinguistics*, *28*(4), 661–677. https://doi.org/10.1017/S014271640707035X

New, B., Pallier, C., Brysbaert, M., & Ferrand, L. (2004). Lexique 2: A new French lexical database. Behavior Research Methods, Instruments, & Computers, 36(3), 516–524. https://doi.org/10.3758/BF03195598

Nicenboim, B., Roettger, T. B., & Vasishth, S. (2018). Using meta-analysis for evidence synthesis: The case of incomplete neutralization in German. Journal of Phonetics, 70, 39–55. https://doi.org/10.1016/j.wocn.2018.06.001

Nosek, B. A., Cohoon, J., Kidwell, M. C., & Spies, J. R. (2015). Estimating the reproducibility of Psychological Science. https://doi.org/10.1126/science.aac4716

Paivio, A. (1971). Imagery and verbal processes. New York: Holt, Rinehart, and Winston.

Piai, V., & Knight, R. T. (2018). Lexical selection with competing distractors: Evidence from left temporal lobe lesions. Psychonomic Bulletin & Review, 25(2), 710-717. https://doi.org/10.3758/s13423-017-1301-0.

Piai, V., Riès, S. K., & Swick, D. (2016). Lesions to Lateral Prefrontal Cortex Impair Lexical Interference Control in Word Production. Frontiers in human neuroscience, 9, 721. https://doi.org/10.3389/fnhum.2015.00721

Piai, V., & Roelofs, A. (2013). Working memory capacity and dual-task interference in picture naming. Acta Psychologica, 142(3), 332–342. https://doi.org/10.1016/j.actpsy.2013.01.006

Piai, V., Roelofs, A., Jensen, O., Schoffelen, J.-M., & Bonnefond, M. (2014). Distinct patterns of brain activity characterise lexical activation and competition in spoken word production. PLOS ONE, *9*(2), e88674. https://doi.org/10.1371/journal.pone.0088674

Piai, V., Roelofs, A., & van der Meij, R. (2012). Event-related potentials and oscillatory brain responses associated with semantic and Stroop-like interference effects in overt naming. Brain Research, 1450, 87–101. https://doi.org/10.1016/j.brainres.2012.02.050

Piai, V., Roelofs, A., & Roete, I. (2015). Semantic interference in picture naming during dual-task performance does not vary with reading ability. Quarterly Journal of Experimental Psychology (2006), 68(9), 1758–1768. https://doi.org/10.1080/17470218.2014.985689

Piai, V., Roelofs, A., & Schriefers, H. (2011). Semantic interference in immediate and delayed naming and reading: Attention and task decisions. Journal of Memory and Language, 64(4), 404–423. https://doi.org/10.1016/j.jml.2011.01.004

Piai, V., Roelofs, A., & Schriefers, H. (2012). Distractor strength and selective attention in picture-naming performance. Memory & Cognition, 40(4), 614–627. https://doi.org/10.3758/s13421-011-0171-3

Pinheiro, J., & Bates, D. (2000). Mixed-effects models in S and S-PLUS. Springer-Verlag. https://doi.org/10.1007/b98882

Python, G., Fargier, R., & Laganaro, M. (2018). When Wine and Apple both help the production of Grapes: ERP evidence for post-lexical semantic facilitation in picture naming. Frontiers in Human Neuroscience, 12. https://doi.org/10.3389/fnhum.2018.00136

R Core Team (2018). R: A language and environment for statistical computing. R Foundation for Statistical Computing, Vienna, Austria. https://www.R-project.org/

Rayner, K., & Springer, C. J. (1986). Graphemic and semantic similarity effects in the picture—Word interference task. British Journal of Psychology, 77(2), 207–222. https://doi.org/10.1111/j.2044-8295.1986.tb01995.x

Rizio, A. A., Moyer, K. J., & Diaz, M. T. (2017). Neural evidence for phonologically based language production deficits in older adults: An fMRI investigation of age-related differences in picture-word interference. Brain and Behavior, 7(4), e00660. https://doi.org/10.1002/brb3.660

Rodríguez-Ferreiro, J., Davies, R., & Cuetos, F. (2014). Semantic domain and grammatical class effects in the picture–word interference paradigm. Language, Cognition and Neuroscience, 29(1), 125–135. https://doi.org/10.1080/01690965.2013.788195

Roelofs, A. (1992). A spreading-activation theory of lemma retrieval in speaking. Cognition, 42(1), 107–142. https://doi.org/10.1016/0010-0277(92)90041-F

Roelofs, A. (1993). Testing a non-decompositional theory of lemma retrieval in speaking: retrieval of verbs. Cognition, 47(1), 59-87. https://doi.org/10.1016/0010-0277(93)90062-Z

Roelofs, A. (2001). Set size and repetition matter: Comment on Caramazza and Costa (2000). Cognition, 80(3), 283–290. https://doi.org/10.1016/s0010-0277(01)00134-2

Roelofs, A. (2003). Goal-referenced selection of verbal action: Modeling attentional control in the Stroop task. Psychological Review, 110(1), 88–125. https://doi.org/10.1037/0033-295X.110.1.88

Roelofs, A. (2008). Tracing attention and the activation flow in spoken word planning using eye movements. Journal of Experimental Psychology: Learning, Memory, and Cognition, 34(2), 353–368. https://doi.org/10.1037/0278-7393.34.2.353

Roelofs, A., & Piai, V. (2017). Distributional analysis of semantic interference in picture naming. Quarterly Journal of Experimental Psychology (2006), 70(4), 782–792. https://doi.org/10.1080/17470218.2016.1165264

Rose, S. B., Aristei, S., Melinger, A., & Abdel Rahman, R. (2019). The closer they are, the more they interfere: Semantic similarity of word distractors increases competition in language production. Journal of Experimental Psychology. Learning, Memory, and Cognition, 45(4), 753–763. https://doi.org/10.1037/xlm0000592

Rosinski, R. R. (1977). Picture-word interference is semantically based. Child Development, 48(2), 643–647. https://doi.org/10.2307/1128667

Sailor, K., & Brooks, P. J. (2014). Do part-whole relations produce facilitation in the picture-word interference task? Quarterly Journal of Experimental Psychology (2006), 67(9), 1768–1785. https://doi.org/10.1080/17470218.2013.870589

Scaltritti, M., Navarrete, E., & Peressotti, F. (2015). Distributional analyses in the picture–word interference paradigm: Exploring the semantic interference and the distractor frequency effects. The Quarterly Journal of Experimental Psychology, 68(7), 1348–1369. https://doi.org/10.1080/17470218.2014.981196

Schnur, T. T., & Martin, R. (2012). Semantic picture-word interference is a postperceptual effect. Psychonomic Bulletin & Review, 19(2), 301–308. https://doi.org/10.3758/s13423-011-0190-x

Schriefers, H., Meyer, A. S., & Levelt, W. J. M. (1990). Exploring the time course of lexical access in language production: Picture-word interference studies. Journal of Memory and Language, 29(1), 86–102. https://doi.org/10.1016/0749-596X(90)90011-N

Schriefers, H., & Teruel, E. (2000). Grammatical gender in noun phrase production: The gender interference effect in German. Journal of Experimental Psychology: Learning, Memory, and Cognition, 26(6), 1368–1377. https://doi.org/10.1037/0278-7393.26.6.1368

Schuster, S., Hawelka, S., Hutzler, F., Kronbichler, M., & Richlan, F. (2016). Words in context: The effects of length, frequency, and predictability on brain responses during natural Reading. Cerebral Cortex, 26(10), 3889–3904. https://doi.org/10.1093/cercor/bhw184

Shao, Z., Meyer, A. S., & Roelofs, A. (2013). Selective and nonselective inhibition of competitors in picture naming. Memory & Cognition, 41(8), 1200–1211. https://doi.org/10.3758/s13421-013-0332-7

Shao, Z., Roelofs, A., Martin, R. C., & Meyer, A. S. (2015). Selective inhibition and naming performance in semantic blocking, picture-word interference, and color-word Stroop tasks.

Journal of Experimental Psychology. Learning, Memory, and Cognition, 41(6), 1806–1820. https://doi.org/10.1037/a0039363

Starreveld, P. A., La Heij, W., & Verdonschot, R. (2013). Time course analysis of the effects of distractor frequency and categorical relatedness in picture naming: An evaluation of the response exclusion account. Language and Cognitive Processes, 28(5), 633–654. https://doi.org/10.1080/01690965.2011.608026

Starreveld, P. A., & La Heij, W. (1995). Semantic interference, orthographic facilitation, and their interaction in naming tasks. Journal of Experimental Psychology: Learning, Memory, and Cognition, 21(3), 686–698. https://doi.org/10.1037/0278-7393.21.3.686

Starreveld, P. A., & La Heij, W. (1996). Time-course analysis of semantic and orthographic context effects in picture naming. Journal of Experimental Psychology: Learning, Memory, and Cognition, 22(4), 896–918. https://doi.org/10.1037/0278-7393.22.4.896

Stroop, J. R. (1935). Studies of interference in serial verbal reactions. Journal of Experimental Psychology, 18(6), 643–662. https://doi.org/10.1037/h0054651

Vasishth, S., & Nicenboim, B. (2016). Statistical methods for linguistic research: Foundational ideas – Part I. Language and Linguistics Compass, 10(8), 349–369. https://doi.org/10.1111/lnc3.12201

Vasishth, S., Nicenboim, B., Beckman, M. E., Li, F., & Jong Kong, E. (2018). Bayesian data analysis in the phonetic sciences: A tutorial introduction. Journal of Phonetics, 71, 147–161. https://doi.org/10.1016/j.wocn.2018.07.008.

Vieth, H. E., McMahon, K. L., & de Zubicaray, G. I. (2014). The roles of shared vs. distinctive conceptual features in lexical access. Frontiers in Psychology, 5, 1014. https://doi.org/10.3389/fpsyg.2014.01014

Vigliocco, G., Vinson, D. P., Lewis, W., & Garrett, M. F. (2004). Representing the meanings of object and action words: The featural and unitary semantic space hypothesis. Cognitive Psychology, 48(4), 422–488. https://doi.org/10.1016/j.cogpsych.2003.09.001

Vitkovitch, M., & Tyrrell, L. (1999). The effects of distractor words on naming pictures at the subordinate level. The Quarterly Journal of Experimental Psychology Section A, 52(4), 905–926. https://doi.org/10.1080/713755854

Zhang, Q., Feng, C., Zhu, X., & Wang, C. (2016). Transforming semantic interference into facilitation in a picture–word interference task. Applied Psycholinguistics, 37(5), 1025–1049. https://doi.org/10.1017/S014271641500034X

Zhu, X., Damian, M. F., & Zhang, Q. (2015). Seriality of semantic and phonological processes during overt speech in Mandarin as revealed by event-related brain potentials. Brain and Language, 144, 16–25. https://doi.org/10.1016/j.bandl.2015.03.007

de Zubicaray, G. I., Hansen, S., & McMahon, K. L. (2013). Differential processing of thematic and categorical conceptual relations in spoken word production. Journal of Experimental Psychology: General, 142(1), 131–142. https://doi.org/10.1037/a0028717

de Zubicaray, G. I., McMahon, K. L. (2009). Auditory context effects in picture naming investigated with event-related fMRI. Cognitive, Affective, & Behavioral Neuroscience 9, 260–269. https://doi.org/10.3758/CABN.9.3.260

**Acknowledgements**

The authors would like to thank all the authors who shared their datasets or took the time to dig into their (sometimes very old) folders to try to find them. They are grateful to Markus Damian and two anonymous reviewers for their helpful feedback on a previous version of this article

This research was funded by the Deutsche Forschungsgemeinschaft (DFG, German Research Foundation) – project number 317633480 – SFB 1287, Project B05 (Bürki) and project Q (Vasishth/Engbert).

***Appendix 1**. Publication statistics for papers with the key words "picture-word interference" and "semantic interference" according to Web of Science*

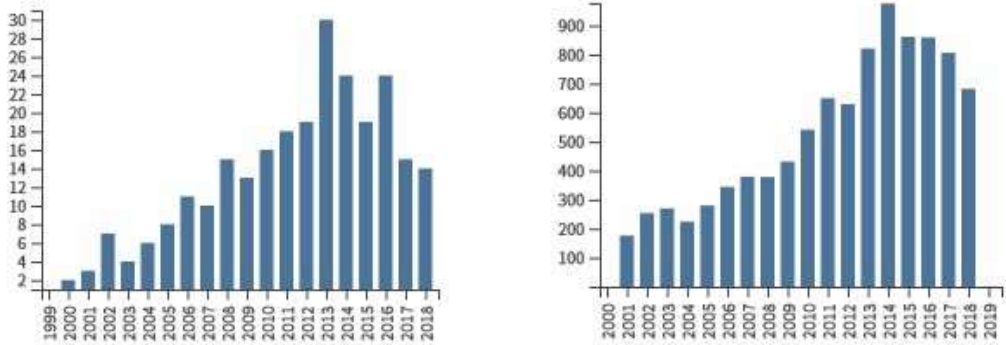

Fig. A.1. Number of publications per year (left) and sum of number of citations per year (right) of articles with the key words "picture-word interference" and "semantic interference".

1 ***Appendix 2.*** *Outcome of studies testing for an interaction between semantic distance and semantic*
2 *relatedness and/or comparing the effects of distractors with different degrees of semantic distance to*
3 *the target word.*

| Study | Outcome | | Metric of semantic distance |
|---|---|---|---|
| | **Interaction** | **Semantic distance effect** | |
| Lupker (1979), Experiment 2 | Interference effect for both, distractors that are also semantic associates (29 ms) and distractors that are not (29 ms) | No difference between members of the same semantic category that are also strong associates (771 ms) and members of the same semantic category with no association (771 ms) | Free association norms |
| Lupker (1979), Experiment 3 | Interference effect for both typical (24 ms) and atypical (30 ms) members of the semantic category | No difference between members of the same semantic category that are typical (734 ms) vs atypical members (740 ms) | Typical members vs. atypical members of semantic category |
| La Heij, Dirkx, and Kramer (1990), Experiment 2, SOA : 0 | Interference effect for weakly associated pairs (17 ms), not significant for strongly associated pairs (-3 ms) | Not tested, between participants | Association norms |
| La Heij, Dirkx, and Kramer (1990), Experiment 2, SOA : -400 | Facilitatory effect (-69 ms) for strongly associated pairs, no significant effect for weakly associated pairs (3 ms) | Not tested, between participants | |
| La Heij, Dirkx, and Kramer (1990), Experiment 3, SOA: +75 | Interference for strongly associated pairs (22ms), not significant for weakly associated pairs (16 ms) | Not tested, between participants | |
| La Heij, Dirkx, and Kramer (1990), Experiment 3, SOA: +150 | Interference for strongly associated pair (29 ms), not significant for weakly associated pairs (9 ms) | Not tested, between participants | |
| Vigliocco et al. (2004), Experiment 3 | Interference for very close, close and moderately close distractors of respectively 29, 15, and 6 ms, with a significant linear trend (as compared to the condition 'far') | Not tested Very close (671 ms), close (657 ms), moderately close (648 ms), unrelated (642 ms) | Vectors of semantic features based on feature descriptions provided by independent group of participants |
| Mahon et al. (2007), Experiment 5 | Interference effect for distractors that are semantically far (37 ms) from the target word, no effect for semantically close distractors (-4 ms) | Semantically close (724 ms) distractors facilitate naming when compared with semantically far distractors (765 ms) | Category members with great vs. small semantic distance selected by experimenter (difference in semantic similarity validated by participants' ratings) |
| Mahon et al. (2007), Experiment 5 b | Interference effect for distractors that are semantically far (37 ms) from the target word no significant effect for semantically close distractors (17 ms). | Semantically close (726 ms) distractors facilitate naming when compared with semantically far distractors (746 ms) | |
| Mahon et al. (2007), Experiment 6 | Not tested | Semantically close (798 ms) distractors facilitate naming when compared with semantically far (814 ms) distractors (16 ms, n.s. by item) | |
| Mahon et al. (2007), Experiment 7, SOA: -160 | Interference effect for semantically far distractors (29 ms) and semantically close distractors (13 ms) | Longer latencies for far (746 ms) than for close distractors (730 ms) | Semantic similarity according to the feature generation norms of Cree and McRae (2003, difference in semantic similarity validated by participants' ratings) |
| Mahon et al. (2007), Experiment 7, | Interference effect for semantically far distractors (16 | No difference (close: 752 ms , far : 748 ms) | |
| SOA: 0 | ms) and semantically close distractors (20 ms) | | |
| Mahon et al. (2007), Experiment 7, SOA: +160 | Interference effect for semantically far distractors (13 ms), no effect for semantically close distractors (3 ms) | No difference (close: 698 ms , far : 708 ms ) | |
| Mahon et al. (2007), Experiment 7b | Interference effect for semantically distractors (18 ms) and semantically close distractors (18 ms) | Naming times are equal for within-category semantically close distractors and within-category semantically far distractors (773 ms) | |
| Hutson and Damian (2014), Experiment 1 | Interference effect for medium (43 ms), close (40 ms) and very close (45) distractors | No difference between medium (846 ms), close (843 ms) and very close (848 ms) pairs | Same as in Vigliocco et al. (2004) |
| Hutson and Damian (2014), Experiment 2 | Interference effect for semantically close (23 ms) and semantically far (19 ms) distractors (n.s. in by-item analysis) | No difference between far (727 ms) and close (731 ms) pairs | Same as in Mahon et al. (2007), Experiment 5 |
| Aristei and Abdel Rahman (2013) | Interference effect (of 6 ms) no longer significant once semantic similarity between distractor and target introduced in statistical model. | Not tested | Semantic similarity between target and distractor pairs rated by independent group of participants |

***Appendix 3.*** *Computation of standard error of the difference between semantically related (r) and unrelated (u) conditions depending on information available.*

|   | Information | Computation |
|---|---|---|
| 1 | Raw dataset | Mixed-effect model testing the contrast of interest with by-item and by-participant random intercepts and slopes |
| 2 | F` | $\sqrt{F'} = t = \dfrac{x_u - x_r}{SEdiff}$ $$SEdiff = \dfrac{x_u - x_r}{\sqrt{F'}}$$ |
| 3 | F1-F2 or t1-t2 (separate by participant and by item analyses) | $F' = (F1 * F2)/(F1 + F2)$ $$SEdiff = \dfrac{x_u - x_r}{\sqrt{F'}}$$ |
| 4 | t (or $(\sqrt{F})$) | $SE_{diff} = \dfrac{x_u - x_r}{t}$ |
| 5 | Standard error of the mean for each condition | Larger of the two SEs (upper bound, following Jägger et al, 2015) |
| 6 | Cohen's d | 1. $SD_{pooled} = \dfrac{x_u - x_r}{d}$ <br> 2. $SE_{diff} = \sqrt{\dfrac{SDpooled^2}{n}}$ |
| 7 | Standard deviation for each condition (r: related, u: unrelated) | 1. $COV(u,r) = rho * SD_u * SD_r$ <br> 2. $VAR(u-r) = \sigma_u^2 + \sigma_r^2 - 2 * COV_{u,r}$ <br> 3. $SE_{diff} = \sqrt{\dfrac{VAR(u-r)}{n}}$ <br><br> rho = 0.983 (correlation between mean for related and mean for unrelated computed across all datasets) |

| 8 | *p* value provided for by-participant and by-item analysis | 1. *F* = quantile 1-p of *F* distribution with df1 and df2<br>2. $SEdiff = abs(\frac{x_u - x_r}{\sqrt{F}})$ |
|---|---|---|
| 9 | *p* value provided for by participant analysis | 1. *F*1 = quantile 1-p of *F* distribution with df1 and df2<br>2. *F*2 = quantile 1-p of *F* distribution with df1 and df2<br>3. $F' = \frac{F1*F2}{F1+F2}$<br>4. $SEdiff = \frac{x_u - x_r}{\sqrt{F'}}$ |
| 10 | F', F1, t1, F1-F2, t1-t2 = ≤ 1 or n.s or *p* > 0.05 | $SEdiff = abs(\frac{x_u - x_r}{1})$ |
| 11 | Separate Standard Errors of the mean for semantically related and unrelated | Larger of the two SEs (upper bound, following Jägger et al, 2015) |

*Appendix 4. Sensitivity analysis for Analysis 3 (Meta-regression with SOA as predictor). The table displays the estimates and credible intervals for the intercept and each contrast, with different priors for the predictor SOA. In a first model we used a normal distribution with a mean of zero and a standard deviation of 200, in a second model we used a uniform distribution bounded between -200 and 200.*

| Prior | Predictor | Estimate (ms) | CrI (95%) |
|---|---|---|---|
| N(0,200) | Intercept (SOA= zero) | 25.4 | 22.0 – 29.2 |
| | Short negative SOA | -7.1 | -13.5 – -0.5 |
| | Short positive SOA | -14.8 | -22.2 – -7.6 |
| | Long negative SOA | -27.5 | -35.0 – -19.7 |
| | Long positive SOA | -24.5 | -34.2 – -14.7 |
| | Between-study SD | 13.0 | 10.9 – 15.4 |
| Uniform(-200,200) | Intercept (SOA= zero) | 25.4 | 21.9 – 29.1 |
| | Short negative SOA | -7.1 | -13.8 – -0.5 |
| | Short positive SOA | -14.7 | -21.8 – -8.0 |
| | Long negative SOA | -27.3 | -34.7 – -19.9 |
| | Long positive SOA | -24.4 | -33.7 – -15.0 |

Between-study SD 13.0 10.9 – 15.5

***Appendix 5.*** *Semantic classification used to compute the number of semantic categories in each study*

| Category | Items |
|---|---|
| Animal | Ant, Ape, Bat, Bear, Beaver, Bee, Bird, Bug, Bumblebee, Butterfly, Camel, Carp, Cat, Caterpillar, Chicken, Cow, Crab, Crocodile, Deer, Dinosaur, Dog, Dolphin, Donkey, Duck, Eagle, Eel, Elephant, Fish, Fly, Fox, Frog, Giraffe, Goat, Goldfish, Gull, Hamster, Hedgehog, Hen, Hippopotamus, Horse, Ibis, Jellyfish, Kangaroo, Ladybug, Leopard, Lion, Mole, Monkey, Mosquito, Moth, Mouse, Octopus, Ostrich, Owl, Panda, Panther, Parrot, Peacock, Penguin, Pig, Pigeon, Python, Rabbit, Rat, Ray, Rhinoceros, Salmon, Scorpio, Sea lion, Seahorse, Seal, Seashell, Shark, Sheep, Snail, Snake, Spider, Spoonbill, Squirrel, Stork, Swan, Swordfish, Tiger, Turkey, Turtle, Vulture, Wasp, Weasel, Whale, Zebra |
| Astronomy | Earth, Moon, Star, Sun |
| Audiovisual objects | Camera, Radio, Television |
| Bags | Backpack, Bag, Sack, Suitcase |
| Body Parts | Arm, Bone, Brain, Chest, Ear, Eye, Face, Finger, Foot, Hair, Hand, Heart, Leg, Lips, Lungs, |

|  | Mouth, Nail, Nose, Shoulder, Skeleton, Tail, Thumb, Tongue |
|---|---|
| Buildings | Bridge, Castle, Church, Dam, Eiffel tower, Factory, Forbidden city, House, Igloo, Jail, Kremlin, Lighthouse, Mecca, Mill, Palace, Prison, Tower, White house, Windmill |
| Clothes | (sleeveless) shirt, Baseball cap, Belt, Beret, Bikini, Boot, Bowler, Bra, Cap, Chapka, Cloak, Coat, Dress, Glove, Hat, Helmet, Hood, Jacket, Pajama, Pants, Pullover, Rice hat, Sandal, Scarf, Shirt, Shoe, Skates, Skirt, Slippers, Sock, Suspenders, Sweat, Sweater, Tie, Trousers, Tuque, Turban, Vest, Waistcoat |
| Color | Black, Purple, Yellow |
| Desk material | Crayons, Eraser, Fountain pen, Pen, Pencil, Ruler |
| Food | Apple, Banana, Borschtsch, Bread, Broccoli, Butter, Cabbage, Cake, Carrot, Celery, Cheese, Cherry, Chestnut, Chickpeas, Corn, Cucumber, Eclair, Eggplant, Grapes, Hamburger, Ice cream, Ice cream cone, Lemon, Lettuce, Oats, Onion, Orange, Peanut, Pear, Pepper, Pineapple, Pizza, Popcorn, Pretzel, Pumpkin, Pumpkins, Rice, Sandwich, Spring roll, Strawberry, Toffee, Tomato, Turnip cabbage, Walnut, Zucchini |
| Flying toys | Balloon, Kite |
| Light devices | Candle, Lamp, Match, Torch |
| Furniture | Air mattress, Bed, Bench, Cabinet, Chair, Couch, Cupboard, Desk, Drawer, Seat, Sofa, Stool, Table, Wardrobe |
| Games | Dart, Dice, Dominos |
| Geometrical shapes | Circle, Square, Triangle |
| Hour devices | Alarm clock, Clock, Watch |
| Musical instruments | Accordion, Cello, Drum, Flute, Guitar, Harp, Organ, Piano, Trumpet, Violin |
| Jewelry | Earring, Necklace, Ring |
| Job | Chef, Clown, Conductor, Cook, Cop, Dentist, Doctor, Lawyer, Nurse, Pilot, Pirate, Policeman, Teacher, Waiter |
| Kitchen | Bottle, Bowl, Chopper, Cork, Cup, Fork, Glass, Jar, Jug, Kettle, Knife, Ladle, Mug, Oven, Pan, Pitcher, Plate, Pot, Spoon, Toaster, Whisk, Wineglass |
| Nationality | American, Chinese, Frenchman, Russian, Saudi |
| Places/landscapes | Bamboo forest, Desert, Hill, Island, Lake, Mountain, Oasis, Prairie, River, Tundra, Vineyard |
| Plants | Acorn, Alga, Basil, Cactus, Dill, Fern, Flower, Leaf, Mushroom, Palm, Palm tree, Plant, Reed, Root, Rose, Tree, Tulip, Vine, Water lily |
| Playground toys | Slide, Swing, Toboggan |
| Reading material | Book, Newspaper, Papyrus |

| Tools/Garden devices | Axe, Bolt, Chisel, Drill, Hammer, Hoe, Ladder, Plane, Pliers, Rake, Saw, Screw, Screwdriver, Shovel, Wheelbarrow |
|---|---|
| Toys | Doll, Spintop, Teddy |
| Vehicle | Airplane, Bicycle, Bike, Boat, Bus, Canoe, Car, Caravan, Carriage, Chair lift, Ferry, Fishing cutter, Fishing vessel, Gondola, Helicopter, Jet, Mobilhome, Motorbike, Motorcycle, Oil tankers, Raft, Sailboat, Scooter, Ship, Submarine, Tractor, Train, Truck, Uboot, Van, Wagon, Yacht |
| Weapons | Arrow, Bomb, Bow, Cannon, Catapult, Dagger, Grenade, Gun, Pistol, Rifle, Rocket, Shield, Spear, Sword, Tank, Truncheon (matraque), Truncheons |
| Undefined | Anchor, Baby, Ball, Banner, Barrel, Basket, Bathtub, Battery, Bell, Bicycle pump, Broom, Buoy, Button, Cage, Calculator, Camera, Capsule, Cave, Cloud, Crane, Cross, Crown, Curtain, Diving mask, Diving suit, Engine, England, Feather, Fence, Fire, Fishhooks, Fishing net, Fishing rod, Flag, Flippers, Floating tyre, Football, Fountain, Hairdryer, Harpoon, Key, Lightning, Lightswitch, Lipstick, Man, Map, Mask, Maze, Medal, Mirror, Mummy, Needle, Paddles, Padlock, Paint, Parachute, Perforator, |
|  | Picnic, Pool, Pram, Rain, Razor, Roller skate, Room, Rope, Rubber, Rudder, Rug, Saddle, Scale, Scissors, Sculpture, Ski, Sky, Snorkel, Soap, Sponge, Stamp, Statue, Surfboard, Swimming cap, Tent, Thermometer, Toll, Toothbrush, Trash can, Vampire, Vase, Well, Wheel, Whip, Whistle, Witch |
| Hair_device | Brush, Comb |
| Smoking object | Cigar, Pipe |
| House_Part | Door, Wall, Window |

**Appendix 6.** *Sensitivity analysis for Analysis 4. In a first model, we changed the priors for the continuous predictors to a normal distribution centered at zero with a standard deviation of 50 and those for the categorical predictors to a normal distribution centered at zero with a standard deviation of 200. In a second model, we changed the priors for the continuous predictors to a uniform distribution bounded between -100 and 100 for continuous predictors and to a uniform distribution bounded between -200 and 200 for categorical predictors.*

| Prior Predictors | | Estimate (ms) | CrI (95%) |
|---|---|---|---|
| N(0,50) for continuous predictors, N(0,200 for categorical predictors) | Intercept | 20.1 | 13.5 – 27.0 |
| | SOA (centered) | -0.03 | -0.08 – 0.02 |
| | Naming times unrelated (centered) | 0.09 | 0.04 – 0.1 |
| | Familiarization | -2.9 | -15.4 – 9.9 |
| | Number of categories (centered) | 0.08 | -0.6 – 0.7 |
| | Number of repetitions (centered) | -0.2 | -0.5 – 0.09 |
| | Set size (centered) | -0.04 | -0.3 – 0.2 |
| | Response set | -2.2 | -10.9 – 6.6 |
| | SOA * RT unrelated | -0.0003 | -0.0008 - 0.0002 |
| | Intercept | 20.0 | 13.4 – 26.7 |
| Uniform(-100,100) for continuous predictors, Uniform(-200,200 for categorical predictors) | SOA (centered) | -0.03 | -0.08 – 0.02 |
| | Naming times unrelated (centered) | 0.09 | 0.04 – 0.13 |
| | Familiarization | -3.0 | -14.9 – 9.7 |
| | Number of categories (centered) | 0.08 | -0.6 – 0.8 |
| | Number of repetitions (centered) | -0.2 | -0.5 – 0.09 |
| | Set size (centered) | -0.05 | -0.3 – 0.2 |
| | Response set | -2.3 | -10.6 – 6.3 |
| | SOA * RT unrelated | -0.0003 | -0.0008 – 0.0003 |

***Appendix 7.*** *Results of sensitivity analyses for Analysis 5 (meta-analyses of the interactions between semantic relatedness and distractor frequency, distractor length, and distractor picturability). Each meta-analysis was run with two different priors for the intercept, a normal distribution with a mean of 0 and standard deviation of 200, and a Uniform distribution bounded between -200 and 200.*

| *Prior Predictors* | *Predictor* | *Estimate (ms)* | *CrI (95%)* |
|---|---|---|---|
| N(0,200) | Distractor frequency * relatedness | -0.8 | -3.4 – 1.8 |
|  | Distractor length * relatedness | 2.7 | -0.06 – 5.5 |
|  | Distractor picturability * relatedness | -0.1 | -7.6 – 6.1 |
| Uniform(-200,200) | Distractor frequency * relatedness | -0.8 | -3.3 – 1.8 |
|  | Distractor length * relatedness | 2.7 | 0.04 – 5.4 |
|  | Distractor picturability * relatedness | -0.3 | -7.3 – 5.8 |